\documentclass[%
 reprint,
 amsmath,amssymb,
 aps,
pre,
]{revtex4-2}

\usepackage{graphicx}
\usepackage{dcolumn}
\usepackage{bm}
\usepackage{xcolor}

\usepackage{caption}
\usepackage{bbold}

\begin{document}

\preprint{APS/123-QED}

\title{Eigenvalue Repulsion and Eigenfunction Localization in Sparse Non-Hermitian Random Matrices}

\author{Grace H. Zhang}
\affiliation{Department of Physics, Harvard University, Cambridge, MA 02138, USA.}
\author{David R. Nelson}
\affiliation{Department of Physics, Harvard University, Cambridge, MA 02138, USA.}%

\date{\today}

\begin{abstract}
Complex networks with directed, local interactions are ubiquitous in nature, and often occur with probabilistic connections due to both intrinsic stochasticity and disordered environments. Sparse non-Hermitian random matrices arise naturally in this context, and are key to describing statistical properties of the non-equilibrium dynamics that emerges from interactions within the network structure. Here, we study one-dimensional (1d) spatial structures and focus on sparse non-Hermitian random matrices in the spirit of tight-binding models in solid state physics. We first investigate two-point eigenvalue correlations in the complex plane for sparse non-Hermitian random matrices using methods developed for the statistical mechanics of inhomogeneous 2d interacting particles. We find that eigenvalue repulsion in the complex plane directly correlates with eigenvector delocalization. In addition, for 1d chains and rings with both disordered nearest neighbor connections and self-interactions, the self-interaction disorder tends to de-correlate eigenvalues and localize eigenvectors more than simple hopping disorder. However, remarkable resistance to eigenvector localization by disorder is provided by large cycles, such as those embodied in 1d periodic boundary conditions under strong directional bias. The directional bias also spatially separates the left and right eigenvectors, leading to interesting dynamics in excitation and response. These phenomena have important implications for asymmetric random networks and highlight a need for mathematical tools to describe and understand them analytically. 
\end{abstract}

\maketitle


\section{Introduction} 
First suggested for the heavy nuclei problem in the mid-twentieth century~\cite{mehta2004random}, random matrix theory has been an important, constantly evolving tool in the studies of large systems with otherwise intractable numbers of degrees of freedom. The first random matrix ensembles, motivated by applications to quantum many-body systems, imposed Hermitian symmetry and all-to-all interactions. However, the past two decades have seen a surge of theoretical and experimental progress identifying and understanding the structure and dynamics of a much wider variety of real-world complex systems. Because interactions within these systems often have a directional bias and depend on spatial or functional locality, their representations as networks and graphs require matrices that are both asymmetric and sparse~\cite{Dorogovtsev2003,dorogovtsev2013evolution,newman2010networks,barabasi2016network,Barrat2008}. 

The spectral characteristics of sparse non-Hermitian random matrices provide information on the stability, susceptibility to perturbations, and synchronization of biological networks~\cite{Rogers2009, Neri2012, Allesina, Barrat2008, Rajan2006,Ahmadian2015,Aljadeff2016, pecora1998master}, guide the construction of practical methods such as graph partitioning and community detection~\cite{krzakala2013spectral,bordenave2015non}, and help evaluate search algorithms~\cite{langville2011google,ermann2015google,noh2004random}.
In addition, advancements in technology in recent years have mapped out the connectivity of large biological systems, such as neural and gene regulatory networks \cite{Kleinfeld2011,Ko2011,Karlebach2008,Chung2013}. These developments motivate the study of sparse random matrices with spatial structure. For example, in neural networks, the anatomical or functional distance between neurons or neural clusters significantly affects their connection probabilities~\cite{Aljadeff2016,Tanaka2018}. Likewise, layered or recurrent architectures of artificial neural networks can also be trained by exploiting existing knowledge about the network structure. 

Unfortunately, classic random matrix theory tools, originally developed for symmetric matrices with dense connectivity~\cite{Ginibre1965,Forrester2007,Akemann2007,Zabrodin2006}, are not always directly applicable to these problems. Thus, a better understanding of sparse non-Hermitian random matrices not only provides rich opportunities for analyzing and constructing complex networks, but also opens new doors in mathematics. Recent theoretical progress includes analytical formulations for the spectral distribution and its support, as well as statistics of eigenvalue outliers and their corresponding eigenvector probability distribution~\cite{Metz2019}. However, one key spectral observable that has received less attention is the two-point eigenvalue correlation, which controls the interplay between eigenmodes central to the behavioral response to external perturbations and small fluctuations. 

In this work, we uncover connections between two-point eigenvalue correlations and the localization of eigenvectors of structured sparse non-Hermitian random matrices. We focus here on matrices with the structure of one-dimensional (1d) tight-binding models in solid state physics~\cite{ashcroft1976solid}, which arise naturally in, e.g., ring attractor neural nets~\cite{Kim2017, Tanaka2018} and in more highly connected networks that nevertheless contain a spatial scale over which the connections fall off~\cite{Amir2016}. We also focus on the effect of random self-interactions on the eigenvalues and eigenvectors, for both undirected and directed ring networks. 
\subsection{Structured sparse non-Hermitian matrices and neural networks}
In general, statistics about the eigenspectra and eigenfunctions are critical for understanding the dynamics of any system in a linear regime that can be described by a coupled system of differential equations, which we'll represent by a random matrix ${\bf M}$. When only a dilute concentration of weakly localized states are activated, the eigenfunctions can also be useful for describing the nonlinear dynamics~\cite{shnerb1998winding}. Of course, level statistics have different physical meanings depending on the specific type of system being modeled. 
In this paper, for concreteness, we will motivate our results in terms of a continuous time recurrent neural network (CTRNN). If $\bf M$ represents the connectivity matrix of the neurons, the full nonlinear firing rate model is,
\begin{equation} \label{eq:neural}
\tau \frac { d r _ { i } ( t ) } { d t } = - r _ { i } ( t ) + f \left[ h _ { i } ( t ) + \sum _ { j } M _ { i j } r _ { j } ( t )  \right],
\end{equation}
where $r_i$ is the deviation from the background firing rate of the neuron on the $i$-th site, $\tau$ is the relaxation time scale for the local firing rate, $h_i$ is the external input to the $i$-th neuron (from, say, a sensory system), $M_{ij}$ describes what is in general an asymmetric connectivity matrix from neuron $j$ to neuron $i$, and $f [ \cdot ]$ is a nonlinear activation function (often with a sigmoidal shape). 

When neural activities are not near saturation, it is convenient to take an activation function of the ``threshold linear'' form, $f ( x ) = ( x + 1 ) \Theta ( x + 1 ) \equiv [ x + 1 ] _ { + }$ ~\cite{Kim2017, dayan2001theoretical}.  When the stimuli exceeds the threshold ($x +1>0$) so as to trigger a response, the firing model describes a linear recurrent neural network~\cite{Amir2016, Tanaka2018},
\begin{equation}\label{eq:neural_linear}
\tau \frac { d r _ { i } ( t ) } { d t } = - r _ { i } ( t ) + h _ { i } ( t ) + \sum _ { j } M _ { i j } r _ { j } ( t ) ,
\end{equation}
where the Heavyside function $\Theta$ is implied and the offset can be taken to be 0 without loss of generality by redefining $h_i(t) \rightarrow h_i(t) + 1$. Such a linearized model is capable of both selective amplification and input integration~\cite{Tanaka2018}. Each eigenmode of the matrix $\bf M$ corresponds to a different neural firing pattern; the corresponding complex eigenvalues indicate the frequency and growth/decay of the firing pattern, while the spatial support of the eigenvector indicates the spatial distribution of the active neural cluster. In the simplest models, the eigenfunction corresponding to the eigenvalue with the largest real part dominates the sustained activity. More generally, the superposition of firing patterns corresponding to nearby eigenvalues in spectral space typically controls the information transfer and computation carried out by the neural network in response to various external stimuli. 

\begin{figure}[tb]
\includegraphics[width=1\columnwidth]{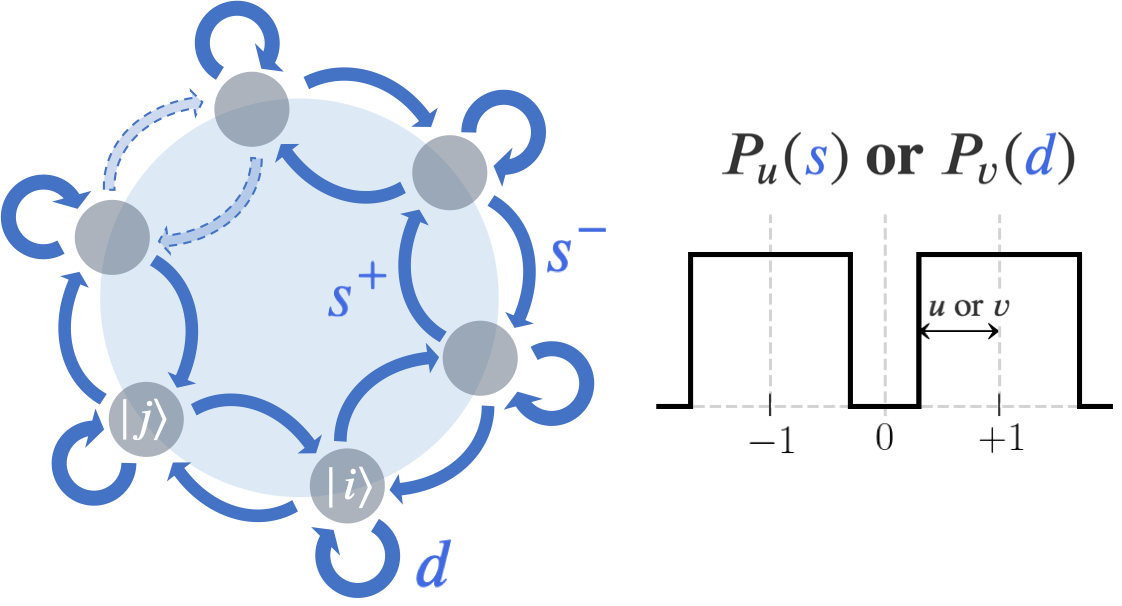}
\caption{\label{fig:1_schm} Left: Schematic representation of Eq.~(\ref{eq:M}) studied in this work. The parameters $s^+$, $s^-$, and $d$  denote site-specific random variables whose distributions describe the randomness of the counterclockwise connections, clockwise connections, and self-interactions, respectively. Right: The form of the independent probability distributions for $s^+$, $s^-$ ($P_u(s)$) and $d$ ($P_v(d)$), where the half-box widths $u$ and $v$ control the ratio of the variance of the connectivity strengths to their mean magnitude.}
\end{figure}

\subsection{Random matrix models}
We study one-dimensional networks, whose interactions are dominated by local spatial couplings. As shown schematically in Fig.~\ref{fig:1_schm}, this connection scheme corresponds to a banded matrix, familiar in condensed matter physics as a tight-binding model, written here in a compact Dirac bra-ket notation as
\small{
\begin{eqnarray} \label{eq:M}
{\bf M} = \sum _ { j = 1 } ^ { N } \left[ s _ { j } ^ { + } e^g | j + 1 \rangle\langle j | +  s _{ j } ^ { - } e^{-g} | j \rangle \langle j + 1| + d_j \epsilon | j \rangle \langle j | \right],
\end{eqnarray}} \normalsize where $s_j^+$, $s_j^-$, and $d_j$ are random variables that can take on both positive and negative values (thus allowing for both excitatory and inhibitory connections), $g$ controls a potential asymmetry of the hopping directional bias ($g>0$ indicates stronger connections in a counterclockwise direction for ring geometries), and $\epsilon$ controls the strength of the random self interactions relative to the random nearest-neighbor connections. Following previous works~\cite{Vogels2005}, we assume balanced inhibition and excitation and investigate symmetric bimodal double-box probability distributions centered on $\pm 1$, for both the diagonal and off diagonal randomness, with box half-widths $u$ and $v$ as tuning parameters (Fig.~\ref{fig:1_schm}). As presented in Eq.~(\ref{eq:M}), this class of models appears to violate Dale's law (connections originating from the same neuron must be all excitatory or all inhibitory~\cite{Rajan2006}). However, as shown in Ref.~\cite{Amir2016}, the spectra for Eq.~(\ref{eq:M}) are identical with those of related models that do obey the constraint. Moreover, if each site in Eq.~(\ref{eq:M}) is regarded as a coarse-grained representation of a cluster of neurons, we expect that the same site can exhibit both excitatory and inhibitory characteristics. 

\begin{figure}[h!]
\includegraphics[width=1\columnwidth]{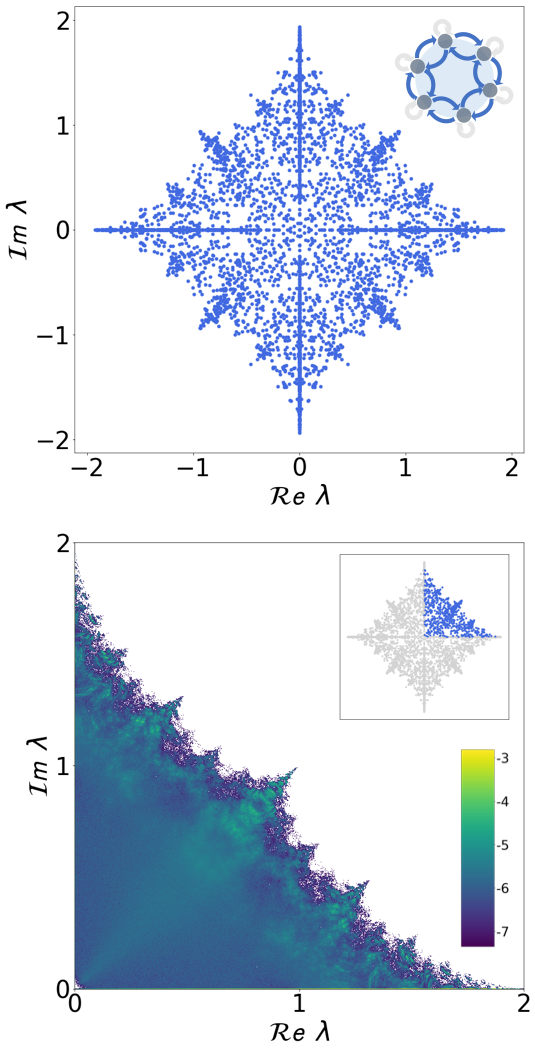}
\caption{\label{fig:2_frac} \linespread{1}\selectfont{Top: Spectrum from a single $N=5000$ realization of the random sign model (Eq.~(\ref{eq:M}) with random sign connection probability distribution $P_{u=0}(s)$, and no self-interaction and directional bias ($\epsilon =g= 0$). Bottom: Spectra in the first quadrant of the complex plane, averaged over $18000$ realizations of $N=5000$ matrices. Color indicates the natural logarithm of the spectral density. Calculation via the box-counting algorithm~\cite{feder2013fractals} gives its boundary fractal dimension to be $D_{bound} = 1.086 \pm 0.004$ and its area fractal dimension as $D_{area} = 1.912 \pm 0.003$.} }
\end{figure}

In Sec.~\ref{sec:repulsion} to Sec.~\ref{sec:self}, we study the spectra of Eq.~(\ref{eq:M}) both with and without disordered self-interactions ($\epsilon > 0$ and $\epsilon = 0$) but no directional bias ($g = 0$). An important parameter in the problem then becomes the ratio $\epsilon$ of the self-interaction strength to the neighboring interaction strength. In the limit of $\epsilon = 0$ with zero variance in the magnitude of the hopping interaction $P_{u=0}(s)$, Eq.~(\ref{eq:M}) is the random sign model, first proposed in Ref.~\cite{Feinberg1999}. We find that its spectrum (shown in Fig.~\ref{fig:2_frac}) is a fractal and, via the box-counting algorithm~\cite{feder2013fractals}, calculate its boundary fractal dimension to be $D_{bound} = 1.086 \pm 0.004$ and its area fractal dimension to be $D_{area} = 1.912 \pm 0.003$. (For comparison, the Hausdorff dimension of the Julia boundary set for $f(z) = z^2 + \frac{1}{4}$ and the Sierpinski carpet are 1.082 and 1.8928, respectively~\cite{feder2013fractals}.)

In Sec.~\ref{sec:oriented}, we examine Eq.~(\ref{eq:M}) in the case of strong directional bias $g \gg 1$ both with and without periodic boundary conditions, $|n \rangle = | N + n \rangle$. In particular, we study a ``one-way'' model, such that counterclockwise interactions on the subdiagonal vanish. We find particularly striking spectra in the limit when the strengths of the diagonal and superdiagonal randomness are equal, for which eigenvalues condense onto the infinity symbol (lemniscate) curve in the complex plane, along which there is a continuous variation in the spatial extent of both the left and right eigenfunctions. 

\subsection{Summary of main results}
This paper focuses on three main themes: 1) eigenvalue repulsion in the complex plane and how it correlates with eigenvalue delocalization in 1d non-Hermitian random matrices; 2) what happens when random self-interactions are added to random nearest-neighbor interactions in 1d tight-binding random matrices; and 3) the effect of directional bias on the localization and spatial separation of the left and right eigenvectors. 

We find that significant eigenvalue repulsion in the complex plane only occurs in the presence of eigenfunction delocalization, and vice versa. Similar results were obtained for localized and extended states in Hermitian tight binding models with diagonal disorder in both three and two dimensions by Shklovskii et al.~\cite{ShklovskiiPRB}. We demonstrate this remarkable correlation numerically for 1d non-Hermitian matrices described by Eq.~(\ref{eq:M}) with neither self-interaction nor directional bias in Sec.~\ref{sec:repulsion} and Sec.~\ref{sec:repulsion_localization}, but emphasize that similar results are obtained for all 1d non-Hermitian random matrices we have examined. In Sec.~\ref{sec:repulsion} and {Appendix~\ref{app:1_gr}}, we describe the procedure used for extraction of the local pair correlation function, whose behavior as a function of the eigenvalue separation depends on the region of the complex spectrum being sampled. In Sec.~\ref{sec:repulsion_localization}, we discuss the relation between the size of the eigenvalue correlation hole in the complex plane and the eigenvector localization length. 

Second, we find that adding disordered self-interactions reduces eigenvalue correlations and enhances eigenvector localization more than random nearest-neighbor connections alone in Eq.~(\ref{eq:M}). We study this phenomenon for systems both with and without directional bias, and with and without periodic boundary conditions. Along the way, we make several interesting observations: In Sec.~\ref{sec:self}, upon adding disordered self-interactions to zero directional bias random nearest neighbor connections, we observe the formation of intricate spectral horns in the complex plane, which elucidates the nature of level mixing under increasing inter-site interactions. First order perturbation theory about basis sets focused on collections of self-inhibiting and self-exciting nodes successfully captures eigenvalue spreading in the limit of weak interactions, and we observe a two-dimensional analog of {electronic band} structure in the complex plane for this disordered system. 

Finally, in Sec.~\ref{sec:oriented}, we study the interplay between directional bias and boundary effects. The large cycle embodied in a ring with periodic boundary conditions, coupled with strong directional bias, leads to eigenvalues confined to a collection of one-dimensional spectral curves with a nontrivial continuous spectral flow connected with eigenvector localization. We identify the spectral curves with equipotential surfaces, generated by charges placed in the complex plane according the probability distribution of the diagonal coefficients. Interestingly, nonzero directional bias leads to the spatial separation of left and right eigenvectors with the same eigenvalue. The formal and physical implications of the asymmetry between left and right eigenvectors is explored in Appendix \ref{app:3_leftright}. 

This paper concludes with a discussion of the physical implications of these results as well as open mathematical questions. 

\section{Eigenvalue correlations} \label{sec:repulsion}

\subsection{Random matrix eigenliquid and the statistical mechanics of interacting particles}

\begin{figure}[h!]
\includegraphics[width=1\columnwidth]{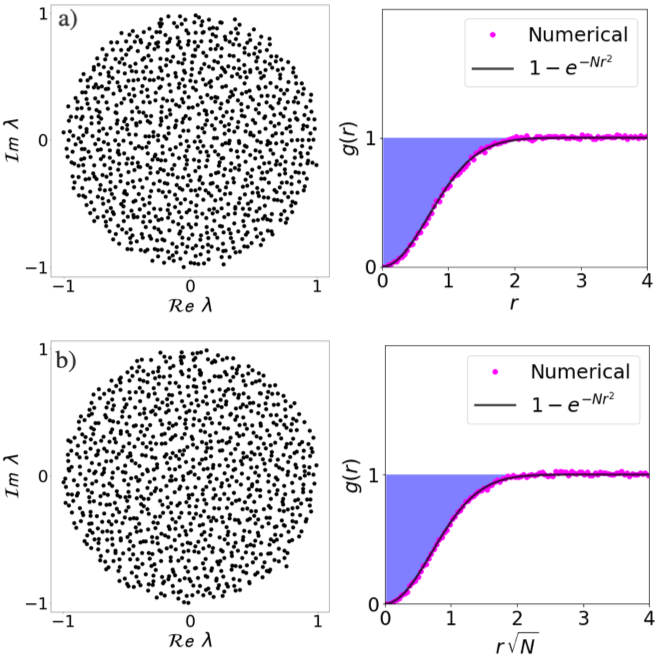}
\caption{\label{fig:3_ginibre}  \linespread{1}\selectfont{(a) Left: spectrum of a single $N=1000$ realization of the complex Ginibre ensemble, where the real and imaginary parts of every matrix element are independently drawn from a Gaussian distribution with mean 0 and variance $1/2 N$.  Right: pair correlation function $g(r)$, where $\vec r = \left (\mathrm{Re}(\lambda-\lambda'), \mathrm{Im} (\lambda - \lambda') \right)$ is the separation between two eigenvalues $\lambda$ and $\lambda'$ on the complex plane, numerically extracted from and averaged over 20 realizations of $N=5000$ matrices from the Ginibre ensemble. Blue shading describes a correlation hole between $g(r)$ and $1$ when $g(r) < 1$; red shading highlights the small region between $g(r)$ and $1$ where $g(r) > 1$. The radial distribution function $g(r)$ vanishes quadratically as $r \rightarrow 0$, indicating logarithmic inter-particle repulsion at short distances with $\beta = 2$, and eventually approaches 1 as $r$ increases, indicating the decay of correlations at large separation distances. The black dashed line plots the analytical expression obtained from direct mapping to a one-component two-dimensional (2d) plasma, as shown in Ref.~\cite{Forrester10log-gasesand}. (b) Left: spectrum of a single $N=1000$ realization of a random matrix where the real and imaginary parts of every matrix element is independently drawn from a uniform distribution with mean 0 and variance $1/2 N$. The spectrum looks qualitatively similar to that of the Ginibre ensemble. Right: pair correlation function $g(r)$, averaged over 20 realizations of $N=5000$ random matrices draw from the uniform distribution. The $g(r)$ curve follows the same functional form as that of the Ginibre ensemble in (a), illustrating the universality of the correlations embodied in Eq.~(\ref{eq:Boltzmann}).}}
\end{figure}

To identify two-point eigenvalue correlations, we treat the eigenvalues as interacting particles in the complex plane and utilize concepts from statistical mechanics. In this section, we outline and demonstrate this method using the complex Ginibre ensemble. The reader is referred to Appendix \ref{app:1_gr} for more details.

The complex Ginibre ensemble~\cite{Ginibre1965} consists of random matrices whose every element has real and imaginary parts separately drawn from independent Gaussian distributions ($M_{jk} = M_{x, jk} + i M_{y, jk}$ where $P(M_{x/y, jk}) \sim \exp(-M_{x/y, jk}^2/2)$). Its eigenvalue joint probability distribution function (JPDF) is known analytically:
{\small{
\begin{eqnarray}\label{eq:Boltzmann}
P\left (\{\lambda\} \right) \sim \exp \left[ -\beta N \left( \sum_{i=1}^N \frac{1}{2} |\lambda_i|^2 -\frac{1}{2 N} \sum_{i \neq j} \ln | \lambda_i - \lambda_j | \right) \right],
\end{eqnarray}}}where the inverse temperature $\beta = 2$ and the eigenvalues have been scaled as $\lambda_i \rightarrow \lambda_i/\sqrt{2 N}$. From Eq.~(\ref{eq:Boltzmann}), it is apparent that the eigenvalue distribution map exactly onto a 2d Coulomb gas under a central harmonic potential. The resulting spectrum follows a ``Circular Law'', in the sense that eigenvalues are uniformly distributed inside a unit circle in the complex plane, with the fraction of eigenvalues lying outside the circle vanishing in the limit $N \rightarrow \infty $ (see Fig.~\ref{fig:3_ginibre}a) \cite{Forrester10log-gasesand}. As also shown in Fig.~\ref{fig:3_ginibre}b, similar results are obtained when the Gaussian distribution is replaced by a box distribution, illustrating that the results for the Ginibre ensemble are universal for a large collection of random matrices with all-to-all connectivities~\cite{Tao2010}. 

Within the uniform disk, however, Eq.~(\ref{eq:Boltzmann}) predicts that eigenvalues experience logarithmic inter-particle repulsion. The key quantity used to characterize correlations among interacting particles in equilibrium statistical mechanics is the radial distribution function (or pair correlation function) $g(r)$, defined in two dimensions as~\cite{mcquarrie2000statistical}
\begin{eqnarray}
\rho g(r) = \rho(r) = \frac{1}{N} \left \langle \sum_{i \neq j} \delta( r- |\vec r_i - \vec r_j|) \right \rangle,
\end{eqnarray}
where $\rho = \left \langle \sum_{j} \delta (\vec r - \vec r_j) \right \rangle$ is the average single particle density and the brackets $\langle \cdots \rangle$ denote average over different realizations of the particle positions $\left \{ \vec r_j  \right\}$. For a particular realization of an ensemble, $g(r)$ indicates the probability of finding a second particle a distance $r$ away from some first particle, given that the first particle exists in that realization. 

By generating multiple realizations, and upon identifying a two-dimensional vector $\vec r_n = (\mathrm{Re} \lambda_n, \mathrm{Im} \lambda_n)$ with each complex eigenvalue $\lambda_n$, we can count all eigenvalue pairs within a range of separation distances. After properly normalizing, we obtain numerically the radial distribution function for the Ginibre ensemble. As seen in Fig.~\ref{fig:3_ginibre}a, $g(r)$ for the Ginibre ensemble contains a correlation hole at small $r$, with a size that scales as $1/\sqrt{N}$, the typical separation distance between the rescaled eigenvalues in the complex plane. In fact, $g(r)$ vanishes quadratically as $r \rightarrow 0$, consistent with the logarithmic inter-particle repulsion in Eq.~(\ref{eq:Boltzmann}) with $\beta = 2$. As $r$ increases, $g(r)$ grows and approaches 1, indicating the decay of correlations at long separation distances, at which particles no longer affect each other. These behaviors are consistent with the analytical expression $g(r) = 1 - e^{-N r^2}$, derived from direct mapping onto a one-component 2d plasma~\cite{Forrester10log-gasesand}. Note the qualitatively similar behavior for the box distribution shown in Fig.~\ref{fig:3_ginibre}b, again illustrating the universality of the Ginibre results for large rank random matrices with independent elements selected from two different probability distributions. 

In evaluating the pair correlation function $g(r)$ for the Ginibre ensemble, we were able to assume a uniform ensemble-averaged density ({$\langle \rho(\vec R) \rangle = \rho ~ \forall \vec R \in $} unit disk). However, as seen in Fig.~\ref{fig:2_frac}, the eigenspectra of the random sign model corresponds to a 2d fluid with both anisotropy and an inhomogeneous density $\rho(\vec R)$. In this case, the pair correlation function does not depend only on the distance between two particles, but more generally also on the global coordinates of the two particles $(\vec r_1, \vec r_2)$. Thus, we will now let $g(r) \rightarrow g(r)_{\vec R}$, which describes the probability of finding a particle at $\vec r_2$ a distance $r$ away from some particle at $\vec r_1$, given that there is a particle at $\vec r_1$ and the mean location of the eigenvalue pair is $\vec R  \equiv  (\vec r_1 + \vec r_2 )/2$.  

\subsection{Level repulsion in nearest neighbor hopping models in one dimension}

We first examine a nearest neighbor hopping model in 1d with no directional bias and no self-interaction ($g=\epsilon = 0$ in Eq.~(\ref{eq:M})), while varying the box-width parameter $u$ of the probability distribution for the hopping term $P_u(s)$ (Fig.~\ref{fig:1_schm}) from $u=0$ (random sign model) to $u = 1$ (single box model). The reason for these choices of parameters will become clear in Sec.~\ref{sec:repulsion_localization}; the resulting spectra strongly suggest that our findings apply more generally to a broad class of sparse non-Hermitian random matrices. 

\begin{figure}[h!]
\includegraphics[width=0.9\columnwidth]{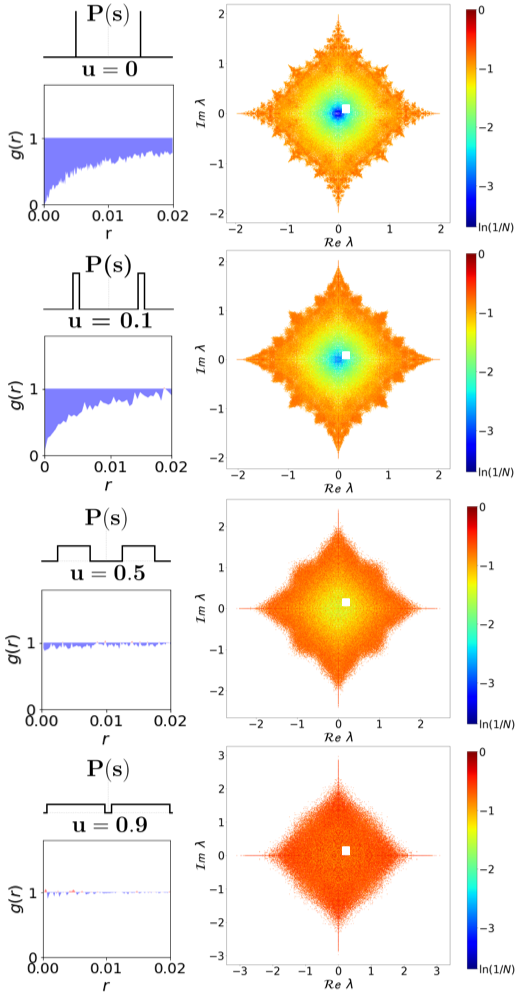}
\caption{\label{fig:4_repulsion} \linespread{1}\selectfont{Eigenvalue pair correlation functions (left) and eigenvalues colored according to the IPR of their corresponding eigenvectors (right) for zero diagonal randomness ($\epsilon=0$) and no directional bias ($g=0$) in Eq.~(\ref{eq:M}). Correlation functions and IPR values are averaged over (710770, 54000, 54000, 423000) and (100, 100, 100, 100) realizations of rank $N=5000$ matrices, respectively, for random hopping strength variance $u = (0.0, 0.1, 0.5, 0.9)$ (see Fig.~\ref{fig:1_schm} for $P_{u}(s)$). For each value of $u$ (increasing from top to bottom), the pair correlation function $g (r)$, with mean eigenvalue pair location $\vec R$ within the small white box on each spectrum, is plotted. For small $u$, the region near the origin of the eigenspectra contains more highly delocalized eigenstates; eigenvalues in that region experience inter-particle repulsion. As $u$ increases, the localization lengths near the origin of the complex plane decrease and the eigenvalues there become de-correlated.}}
\end{figure}

\begin{figure}[h!]
\includegraphics[width=0.88\columnwidth]{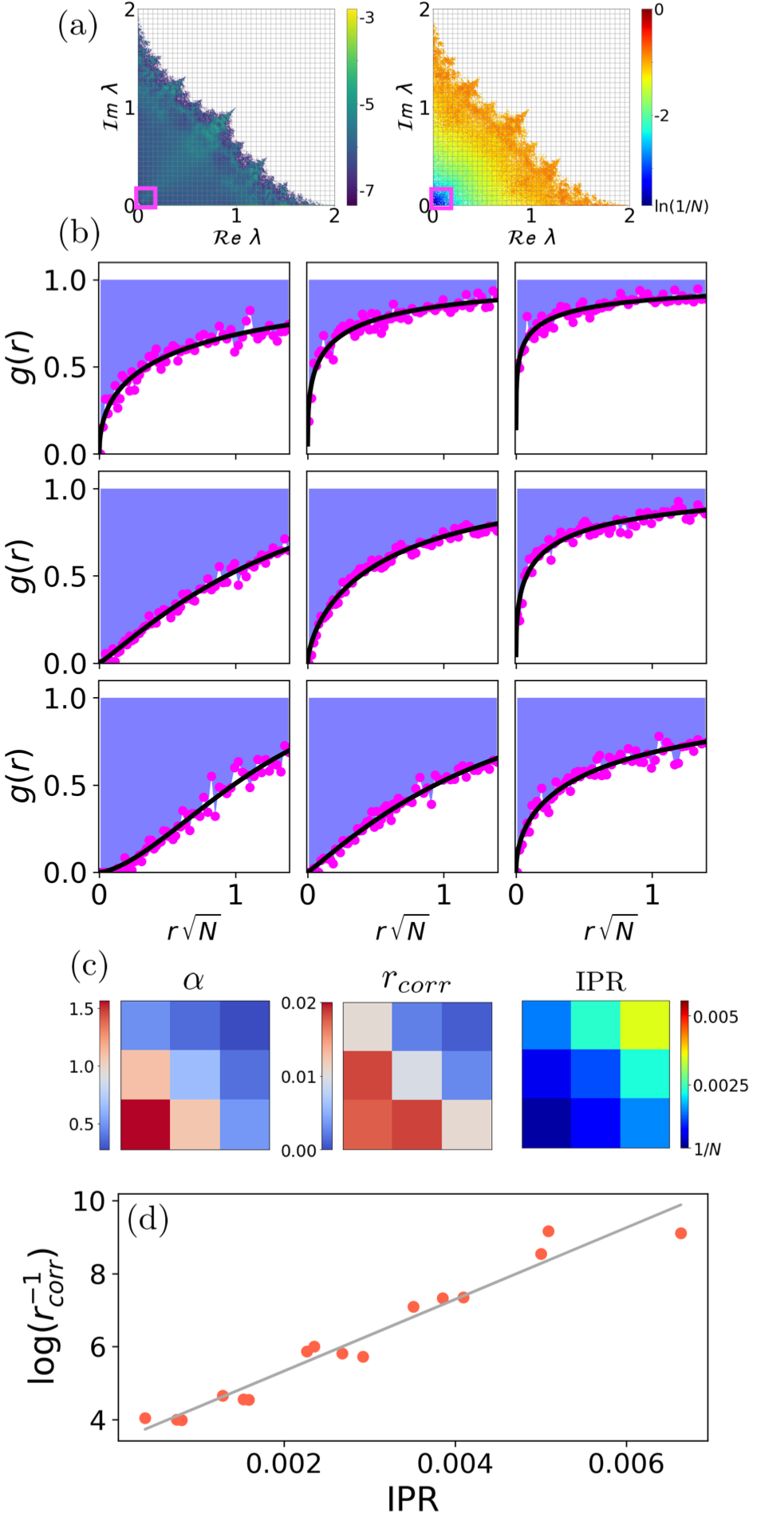}
\caption{\label{fig:3b_gr_fit} (a): Local pair correlation functions $g(r)$ examined for eigenvalue pairs in the 9 ($0.05 \times 0.05$) square grids closest to the origin of the complex plane, enclosed by the magenta box in the $u=0$ random sign spectra [colored by average eigenvalue density (left) and eigenvector IPR (right)]. 
(b): Numerical correlation data of the random sign spectra fit to $ g_s(r) = 1 - \exp \left[ -\left(\frac{r}{r_{corr}} \right)^\alpha \right]$. 
(c): Quantities corresponding to eigenvalues in each of the 9 squares grids. As eigenvalue magnitude increases, the correlation hole width $r_{corr}$ decreases, the exponent $\alpha$ decreases, and the average IPR of the corresponding eigenvectors increases. 
(d): Logarithm of the inverse correlation hole width $r_{corr}^{-1}$ versus IPR, corresponding to eigenvalues in the 16 square grids closest to the origin of the complex plane (including the 9 shown in (b) and (c)). The linear fitting shows exponential dependence, from which Eq.~(\ref{eq:rl}) follows.}
\end{figure}

Due to a singular spike in eigenvalue densities on the real and imaginary axes (see Fig.~\ref{fig:2_frac} and Ref.~\cite{Amir2016}), we extract our local pair correlation function $g(r)_{\vec R}$ (denoted as $g(r)$ henceforth) for the bulk eigenvalues inside quadrants I--IV and along the real and imaginary axes separately. We treat the former as a two-dimensional inhomogeneous fluid and the latter as a one-dimensional inhomogeneous fluid (see Appendix \ref{app:1_gr} for details on numerical methods and normalization procedure). We approximate the 2d eigenfluid inside the various quadrants to be isotropic and justify this approximation in Appendix \ref{app:2_angular} by examining the weak directional variation of $g(r)$ when $r$ is fr from the coordinate axes. Each local pair correlation function $g(r)$ shown in Figs.~\ref{fig:4_repulsion} and \ref{fig:3b_gr_fit} is averaged over a small region of the spectrum within which the eigenvalue density is approximately homogeneous. The right column of Fig.~\ref{fig:4_repulsion} shows the spectra for ${u = (0,~0.1,~0.5,~0.9)}$, averaged over 100 realizations for each value of $u$. As $u$ increases, one can observe that the exact and statistical symmetries of the eigenvalue distribution, as well as its overall diamond-like shape, stay the same \cite{Amir2016}, but details of the fractal edges become smeared out. Our findings for two-point eigenvalue correlations are as follows:

First, eigenvalues at a distance sufficiently far from the origin, and also not too close to the spectral edges so as to experience boundary effects, are uncorrelated (as in an ideal gas) for all values of $u$. However, eigenvalues close to the origin behave differently. For $u = 0$, where the nearest neighbor matrix elements are randomly chosen to be $\pm 1$,  $g(r)$ dips significantly below 1 as $r$ approaches 0 and vanishes for $r = 0$, whereas for large $r$, $g(r)$ grows and plateaus to 1 (Fig.~\ref{fig:3b_gr_fit} and Fig.~\ref{fig:4_repulsion}). This behavior is reminiscent of that of the bulk eigenvalues of the Ginibre ensemble in Fig.~\ref{fig:3_ginibre}. In other words, we discover that for the random sign model, eigenvalues near the origin experience inter-particle repulsion. 

The exact form of the repulsion is different from that of the Ginibre ensemble, as $g(r)$ approaches 0 for small $r$ with a different functional form than the quadratic vanishing we see in Fig.~\ref{fig:3_ginibre} (Sec.~\ref{sec:repulsion_localization}). The left of Fig.~\ref{fig:4_repulsion} shows examples of $g(r)$ for the eigenvalues close to the origin, averaged over (710770, 54000, 54000, 423000) realizations for $u = (0,~0.1,~0.5,~0.9)$, respectively. The regions of the spectra for which we evaluate the local $g(r)$ are indicated by small, white, off-center squares on the right of  Fig.~\ref{fig:4_repulsion}. 
As the box-width $u$ increases, the region near the origin of the complex plane in which eigenvalue pairs experience repulsion with each other shrinks. When $u$ is large enough, the eigenvalues are entirely uncorrelated and the correlations approximate those of an ideal gas everywhere in the spectrum. 

We have applied the same analysis to eigenvalues on the real and imaginary axes, treating them as 1d ensembles. The statistical symmetry of the spectra under $90^\circ$ rotation~\cite{Amir2016} insures identical behavior on these two axes. Since axial eigenvalues near the edge of the spectra exhibit fractal modulations in the eigenvalue density, we examine axial eigenvalues in the region near the origin of the complex plane, where the average eigenvalue density increases linearly along the axis, with increasing distance from the origin~\cite{Amir2016}. The behavior of $g(r)$ for the axial eigenvalues in this region, as a function of probability distribution box-width $u$, is qualitatively consistent with that of the bulk correlations. Specifically, the range and strength of 1d eigenvalue repulsion along the axes, as well as the radial extent along the axes in which eigenvalues experience that repulsion, are largest for $u=0$ and decreases as $u$ increases. The correlations vanish as $u \rightarrow 1$. 

Finally, we also examined the evolution of the eigenspectra under other balanced bimodal distributions, specifically a bimodal Gaussian distribution centered at $\pm1$. The results are qualitatively similar: as the variance of the Gaussian increases, delocalized states and eigenvalue repulsion both disappear. 

\section{Eigenvalue repulsion and eigenfunction delocalization}
\label{sec:repulsion_localization}
What determines the regions in the eigenspectra over which the eigenvalues experience inter-particle repulsion? To better understand our findings in the previous section, we examine the localization properties of the eigenfunctions. 

The main metric we use to characterize the degree of localization of an eigenfunction is the Inverse Participation Ratio (IPR), defined as follows: For the $n$-th eigenvalue $\lambda_n$, the IPR of the right eigenfunction $\psi_n^R$ is
\begin{eqnarray} \label{eq:IPR}
\text{IPR}(\lambda_n) \equiv \left[\frac{\left(\sum_i |\psi_n^R(i)|^2 \right)^2}{\sum_i |\psi_n^R(i)|^4} \right]^{-1},
\end{eqnarray}
where $|\psi_n^R(i)|$ is the amplitude of the $n$-th right eigenfunction at site $i$. Here, we focus on localization properties of the \textit{right} eigenfunctions of the asymmetric random matrices. The left eigenfunctions, as well as the the inner product of the left and right eigenfunctions, behave in a similar fashion for $g=0$  (see Appendix \ref{app:3_leftright}). The IPR varies from being $O(1/N)$ for eigenfunctions spread uniformly across all sites, to $O(1)$ for those localized near a specific site. For each spectrum examined in this work, we have also calculated the Lyapunov exponents and confirmed that they are consistent with the behavior of the IPR.

Heat maps of the IPR for the random hopping eigenspectra with box-widths $u = (0.0, 0.1, 0.5, 0.9)$ are shown on the right of Fig.~\ref{fig:4_repulsion}. For $u=0$,  the localization lengths of the eigenfunctions diverge as their eigenvalues approach the origin, as analyzed in detail in Ref.~\cite{Amir2016}. 
More generally, for small $u$, there is a region near the origin of the eigenspectra that contains rather delocalized eigenstates. Note that as $u$ increases, the region of extended states centered at the origin of the complex plane shrinks and disappears, such that the complex plane is eventually populated entirely by localized eigenstates as $u \rightarrow 1$. 

These findings correlate strongly with our results on eigenvalue repulsion from the previous section: the eigenvalue repulsion near the origin is only present when the more extended eigenvectors are also present. Conversely, when states are highly localized, as near the edge of the spectrum for $u=0$, or everywhere in the complex plane for $u=0.9$, there is no level repulsion and the eigenvalues behave like an ideal gas. We have observed this connection between eigenvalue repulsion and extended eigenstates for \textit{all} non-Hermitian random matrices we have examined. There is no way for highly localized eigenfunctions at very different locations in a one-dimensional lattice to know about each other, so it is plausible that their eigenvalues are uncorrelated. Similar correlations for eigenvalue spacings along the 1d real axis and Anderson localization have been seen in various Hermitian disordered systems by studying the nearest-neighbor spacing distribution (see for example \cite{ShklovskiiPRB,altshuler1986repulsion}).
We conjecture here, for non-Hermitian random matrices with a complex spectrum, that when the eigenfunctions are delocalized, their complex eigenvalues repel each other, and conversely, when eigenvalues repel each other, their eigenfunctions are delocalized.

We can make this connection more precise using the spectra of the $u=0$ random sign model. Fig.~\ref{fig:3b_gr_fit} shows the local correlation functions $g(r)$ in the 9 (0.05 $\times$ 0.05) square grids closest to the origin of the complex plane, enclosed by the magenta box in the spectra shown in the top panel. Motivated by the correlation function of the Ginibre ensemble $g_{G}(r) = 1 - \exp (-Nr^2)$ (Fig.~\ref{fig:3_ginibre}), we fit the numerical correlation data of the spectra for the $u=0$ random sign model to the following function,
\begin{eqnarray} \label{eq:grfit}
g_s(r) = 1 - \exp \left[ -\left(\frac{r}{r_{corr}} \right)^\alpha \right],
\end{eqnarray}
which allows us to extract the width of the correlation hole $r_{corr}$  and the exponent $\alpha$ characterizing the vanishing of the correlations as $r$ goes to 0. As shown in Fig.~\ref{fig:3b_gr_fit}c, as the magnitude of the eigenvalue (i.e. its distance from the origin) increases, the correlation hole $r_{corr}$ decreases (the spatial extent of the inter-particle repulsion shrinks) and the exponent $\alpha$ decreases (the correlation function $g(r)$ approaches 1 more sharply as $r$ increases). Furthermore, the inverse correlation hole width $r_{corr}^{-1}$ appears to depend exponentially on the IPR (Fig.~\ref{fig:3b_gr_fit}d). In one-dimensional systems, the IPR is in fact just the inverse of the localization length $l_{loc}$~\cite{Thouless_1972}. Upon rescaling $\bar r_{corr} = r_{corr}\sqrt{N}$ and $\bar l_{loc} = l_{loc}/N$, where $1/\sqrt{N}$ is the average interparticle spacing of N eigenvalues spanning a 2d complex spectrum of area $O(1)$ and $\bar l_{loc}$ measures the fraction of the $N$-site ring occupied by an eigenfunction, we find that the eigenvalue correlation hole width $\bar r_{corr}$ and the eigenvector localization length $\bar l_{loc}$ are related as follows:
\begin{eqnarray} \label{eq:rl}
\bar r_{corr} = c_1\exp\left(-\frac{c_2}{\bar l_{loc}} \right),
\end{eqnarray}
where $c_1 = 2.4 \pm 0.2$ and $c_2 = 0.196 \pm 0.005$. According to the conjecture embodied in Eq.~(\ref{eq:rl}), $\bar r_{corr}$ vanishes in the limit of strongly localized eigenvectors ($\bar l_{loc} \rightarrow 0$) and increases when eigenvectors become more delocalized ($\bar l_{loc}$ increases) .

The interpretation of this relation between eigenvalue repulsion and eigenvector delocalization for neural networks is as follows: The $n$-th eigenmode of the connectivity matrix ${\bf M}$ in a dynamical model like Eq.~(\ref{eq:neural}) corresponds to a firing pattern $\psi_n(i)$; the pattern is selectively distributed over certain neurons according to the sites $i$ at which the eigenfunction amplitude is nonzero, and these neurons collectively fire, with growth or decay, and oscillations of the firing rates controlled by the complex eigenvalue $\lambda_n$. Given two distinct firing patterns corresponding to two eigenmodes of the connectivity matrix ${\bf M}$, the higher the number of active neurons participating in each of these firing patterns (i.e. the more delocalized the normal modes), the more separated their firing frequencies and growth/decay rates, represented by the eigenvalues in the complex plane.

We also comment on the universality of eigenvalue correlations for large rank random matrices. First, recall that, as illustrated in Fig.~\ref{fig:3_ginibre}, eigenvalue correlations of dense non-Hermitian random matrices are insensitive to changes in the specific shape of the matrix element probability distribution. For example, in the Ginibre ensemble, each matrix element is independently drawn from a Gaussian distribution, yielding the correlations in Fig.~\ref{fig:3_ginibre}a. If the elements were instead drawn from a box distribution with the same mean, the pair correlation function $g(r)$ does not change except up to a possible rescaling of $r$ \cite{tao2011random}, as illustrated in Fig.~\ref{fig:3_ginibre}b. In contrast, the sparse one-dimensional non-Hermitian random matrices studied here are more sensitive to variations in the matrix element probability distribution (see Fig.~\ref{fig:4_repulsion}). The eigenvalue correlations change \textit{qualitatively} with the parameter $u$ in the matrix element probability distribution. Nevertheless, the behavior of the spectra for the large $N$ random one-dimensional hopping models considered so far \textit{are} invariant to changes in boundary conditions (open or periodic). This insensitivity to boundary conditions will be violated dramatically for the models with directional bias examined in Sec.~\ref{sec:oriented} of this paper.

\section{Effects of self interaction} \label{sec:self}

\begin{figure}[h!]
\includegraphics[width=1\columnwidth]{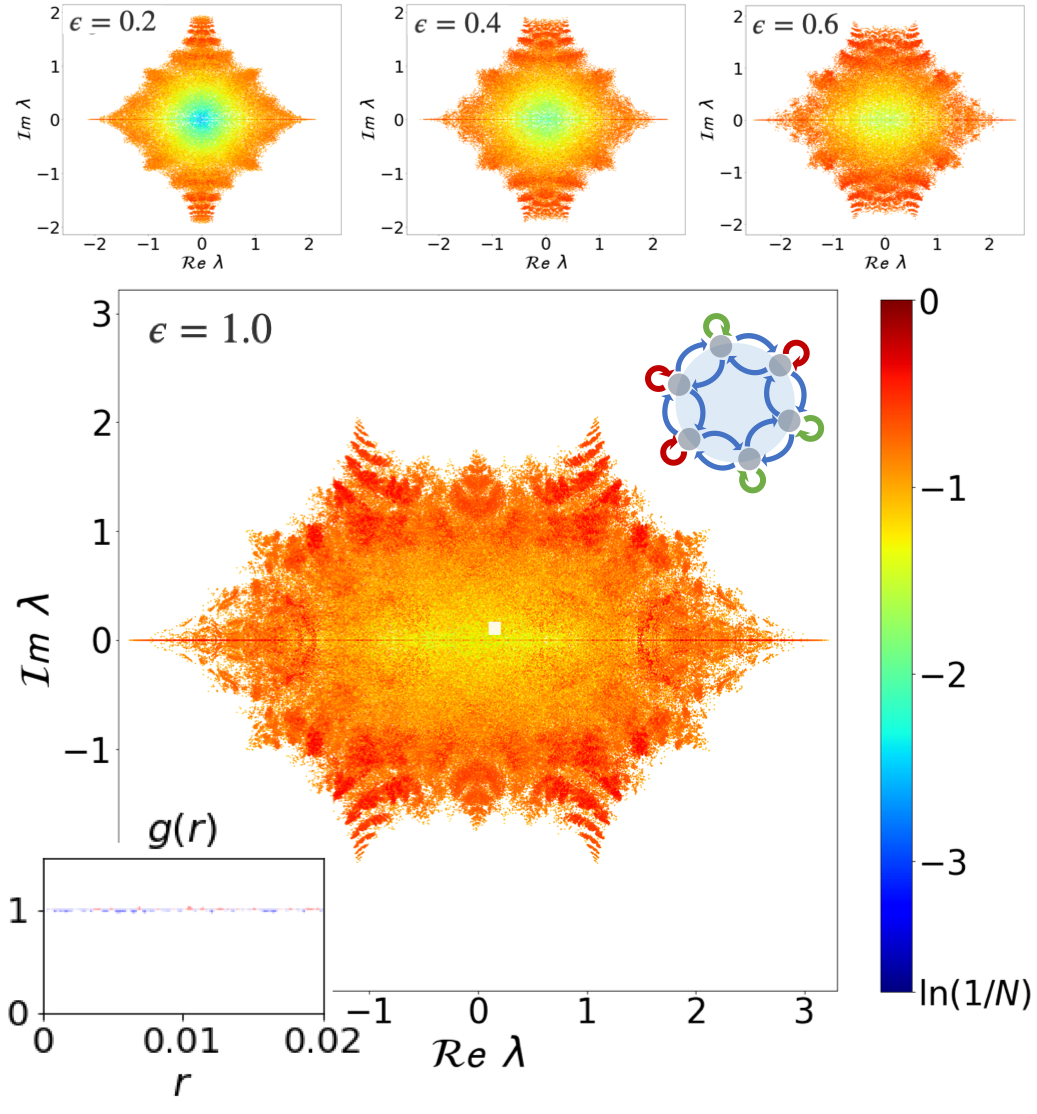}
\caption{\label{fig:5_horn} \linespread{1}\selectfont{Top: Spectra of Eq.~(\ref{eq:M}) with random sign nearest-neighbor coupling probability $P_{u=0}(s)$ and random sign self-interactions distribution $P_{v=0}(d)$ as the self interaction strength $\epsilon$ is tuned from 0 to 1. Bottom: Spectra of Eq.~(\ref{eq:M}) when random self-interaction and random hopping have equal strengths, with $g = 0$ and $\epsilon = 1$. Eigenvalues form four new horn-like boundaries hovering above and below $\pm 1$ along the real axis, which are the values of the diagonal elements $\pm \epsilon$. Compared to the random sign hopping spectrum at the top of Fig.~\ref{fig:4_repulsion}, the addition of random on-site disorder removes both the eigenvalue repulsion and the weakly localized eigenstates near the origin of the complex plane. The lower inset shows the radial distribution function $g(r)$ with mean pair location $\vec R$ in the small white box near the origin, averaged over 9000 realizations of $N=5000$ matrices. The pair correlation function $g(r)$ is a flat line at $1$, showing the eigenvalues behaving like an ideal gas with no correlations.}}
\end{figure}

Generally, network models incorporate self-interactions via nonzero diagonal elements. In biological networks, these feedback effects are referred to as ``self-inhibition'' (or ``self-regulation'') when the diagonal matrix element is negative, and ``self-excitation'' when the diagonal matrix element is positive. In condensed matter physics, such couplings are exemplified by ``onsite-disorder''~\cite{shklovskii2013electronic}. 
In this section, we take $\epsilon > 0$, the strength of the diagonal disorder in Eq.~(\ref{eq:M}), in order to study the effects of self-interactions through probabilistic on-site elements with random signs in the connectivity matrix $\bf M$ of, say, a neural network. In general, one could consider a wide variety of probability distributions for the self-interacting coefficients. We focus our attention here on the case of the random sign distribution $P_{v=0}(d)$ for the diagonal matrix elements (identical in form to the probability distribution $P_{u=0}(s)$ for the random nearest-neighbor connections), because it exhibits important new features and is tractable enough for analysis. 

\subsection{Eigenfunctions in an eigengas generated by diagonal disorder are localized.}
How do disordered self-interactions affect the spectrum and the localization of its eigenfunctions compared to the hopping-only random sign model? To answer this question, we first study Eq.~(\ref{eq:M}), with $g=0$ and strictly bimodal $\pm1$ interactions for both the hopping and diagonal matrix elements, i.e. with probability distributions $P_{u=0}(s)$ and $P_{v=0}(d)$ shown in Fig.~\ref{fig:1_schm}. However, we now vary the relative magnitude of the diagonal disorder by tuning the parameter $\epsilon > 0$. 

Upon turning on the random self-interaction strength $\epsilon$ with diagonal probability distribution $P_{v=0}(d)$ (i.e. the diagonal matrix elements are $\pm \epsilon$, each with probability $1/2$), the nearly extended states shown near the origin at the top panel of Fig.~\ref{fig:4_repulsion} start to disappear.
By the time $\epsilon$ reaches $ 1$, such that the self-interactions have the same level of disorder as the hopping interactions, all eigenstates are strongly localized. The resulting eigenvalue distribution and variation in eigenvector localization, averaged over 200 realizations, are shown in Fig.~\ref{fig:5_horn}. Examination of eigenvalue correlations reveal the radial distribution function $g (r) \approx 1$ for all regions in the complex plane within the spectral support. An example is shown in the inset of Fig.~\ref{fig:5_horn}. In other words, the addition of strong diagonal randomness decorrelates all eigenvalues; the eigenfluid studied in Sec.~\ref{sec:repulsion} has lost its inter-particle interactions and behaves instead as a 2d ideal gas. We performed the same analysis for nonzero distribution box-width $u > 0$, and found the same results qualitatively---eigenvalues do not experience correlations and all eigenvectors are strongly localized. These results are consistent with our findings from Sec.~\ref{sec:repulsion_localization}, and confirms our conjecture that localized eigenfunctions lead to no level repulsion.

\subsection{Spectral horn formation and complex level mixing in the complex plane}

The second intriguing feature of Fig.~\ref{fig:5_horn} is the accumulation of eigenvalues onto a pattern of spectral ``horns'', whose real parts are predominantly at $\pm 1$ on the complex plane. (For large $N$ random matrices with $g=0$, these and other features of the spectra described here and in the following are insensitive to applying open or periodic boundary conditions, as for the random hopping model.)

\begin{figure}[h!]
\includegraphics[width=1\columnwidth]{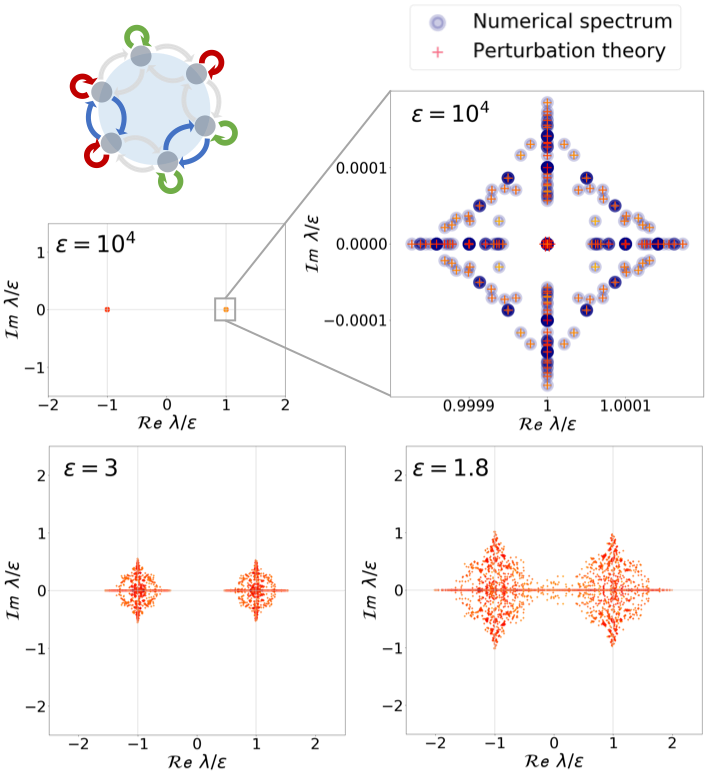}
\caption{\label{fig:6_drand_pert} \linespread{1}\selectfont{Spectra of Eq.~(\ref{eq:M}) with random sign hopping $P_{u=0}(s)$ (i.e. $s_j^+$ and $s_j^- = \pm 1$ with equal probabilities) and random sign self-interactions $P_{v=0}(d)$, where the self-interaction strength is much greater than the hopping interaction strength, $\epsilon \gg 1$. The clustering of eigenvalues around the spectral horns at $\pm\epsilon$ is due to the self-interaction coefficients being distributed at these values. When hopping interactions are turned on, eigenvalues bloom from their degenerate point condensations at $\pm \epsilon$ on the real line into two almost-continuous patches in the complex plane. Top right: First order perturbation theory in $1/\epsilon$ about a system of two disconnected sub-networks, consisting of only self-exciting neurons or only self-inhibiting neurons, captures the start of the spectral blooms at $\epsilon = 10^4$. The neural connections kept in this approximation are shown schematically in the upper left.}}
\end{figure}

It is useful to observe the formation of these spectral horns from a different regime, starting with strong self-interactions and negligible interactions between sites ($\epsilon \rightarrow \infty$). As $\epsilon$ is gradually reduced, the random hopping terms, with probability distribution $P_{u=0}(s)$, become more important, and the eigenspectra evolve as in Fig.~\ref{fig:6_drand_pert}. The spectral horns at $\pm \epsilon$ emerge because the self-interaction coefficients are distributed at $\pm \epsilon$. When hopping interactions are turned on, eigenvalues bloom from their degenerate point condensations on the real line into almost-continuous patches in the complex plane. This phenomenon is the non-Hermitian analog of band theory in quantum condensed matter, where, when isolated identical atoms are brought closer together and begin interacting as they do in dense solids, single-atom energy levels broaden into a continuous electronic band structure. 

In the context of neural networks, the atomic orbitals which lead to discrete energy levels are replaced by self-inhibiting and self-exciting neurons. For large $\epsilon$, if one scales out $\epsilon$, these neurons  at first do not interact with each other, and exhibit locally either purely growing and saturating or purely decaying firing rates. As $\epsilon$ is decreased and connections between neurons become more important, the eigenspectrum expands about $\pm \epsilon$ on the real axis, and the firing patterns each spread out over more participating neurons (marked by the small but noticeable decrease in the IPR for eigenfunctions with eigenvalues close to the origin in Fig.~\ref{fig:5_horn}) and experience a richer set of oscillatory and growth/decay behavior. However, the presence of self-inhibition and self-excitation still dominates the dynamics. For the situation shown in Fig.~\ref{fig:6_drand_pert}, with the spectrum centered on the origin, the eigenmodes separate into a group of mostly growing modes and a group of mostly decaying modes, as indicated by the high density patches of eigenvalues within the spectral horns centered at $\pm \epsilon$. We study this phenomena quantitatively and extract further physical insights using a perturbative approximation in the following section. 

\subsection{Perturbation theory in $1/\epsilon$}
\label{sec:self_no_direction}

Given a connectivity matrix $\mathbf{M}$, one can always reshuffle the basis to restructure $\mathbf{M}$ in terms of the following matrix blocks: the matrix of all connections between neurons that are self-excitatory $\mathbf{M_{+1}}$, the matrix of connections between all neurons that are self-inhibitory $\mathbf{M_{-1}}$ (we explicitly exclude diagonal matrix elements from $\mathbf{M_{+1}}$ and $\mathbf{M_{-1}}$), the matrix of connections from self-exciting neurons to self-inhibiting neurons $\mathbf{C_{+-}}$, and the matrix of connections from self-inhibiting neurons to self-exciting neurons $\mathbf{C_{-+}}$:
\begin{equation}
\mathbf{M} = \epsilon \left[ \begin{matrix}
 \mathbb{1}_{N^+} & 0 \\
 0 			& -\mathbb{1}_{N^-} \end{matrix} \right] + \left[ \begin{matrix}
 \mathbf{M_{+1}} & \mathbf{C_{+-}} \\
 \mathbf{C_{-+}} 			& \mathbf{M_{-1}} \end{matrix} \right],
 \end{equation}
 where $N^+$ and $N^-$ is the number of self-exciting and self-inhibiting neurons, respectively, and $\mathbf{M_{\pm 1}}$ and $C_{\pm \mp}$ all contain elements of order $O(1)$. For $1/\epsilon \ll 1$, one can neglect the off-diagonal matrices  $\mathbf{C_{+-}}$ and  $\mathbf{C_{-+}}$, since their contribution to the eigenvalue vanishes to first order in perturbation theory. $\bf M$ can then be approximated in block diagonal form.
Upon comparing the eigenvalues of Eq.~(\ref{eq:drand_pert}) to those from the exact diagonalization of the full matrix $\bf M$, we see that this perturbative approximation captures the start of the spectral bloom perfectly (top right panel of Fig.~\ref{fig:6_drand_pert}). 

For neural networks, this approximation implies that when the random hopping interaction strengths are weak compared to the self-interactions, the neural network can be approximated by two sub-networks of only self-exciting neurons and only self-inhibiting neurons, and connections between the subnetworks can be neglected (see schematic at top left of Fig.~\ref{fig:6_drand_pert}). 

\section{Coupling self-interactions with strong directional bias} \label{sec:oriented}

Thus far, we have examined Eq.~(\ref{eq:M}) for different scenarios all without directional bias (i.e. $g=0$ in Eq.~(\ref{eq:M})). The parameter $g$ controls the directionality, or orientability, of the network, and has been examined in the context of random hopping models motivated by vortex physics in high temperature superconductors and neural networks \cite{Hatano1997,Amir2016}. In this section, we study the infinite bias limit of an oriented network (the ``one-way'' model of Feinberg and Zee~\cite{Feinberg1999}) with the addition of random self-interactions. 

Mathematically, an ``oriented graph'' is a graph where there can only exist one directed connection between any pair of nodes. In the language of the matrix model in Eq.~\ref{eq:M}, only one of $M_{ij}$ and $M_{ji}$ can be nonzero. In the case of a 1d ring network, the most interesting case is when all connections are pointed in the same direction. If the directionality is counter-clockwise, $\bf M$ has only a nonzero superdiagonal and a zero subdiagonal, as well as a nonzero corner matrix element in the lower left. (Clockwise directionality leads to similar structure on the subdiagonal and in the upper right corner.) Such systems can be understood via the recursion relation for the cofactor expansion of a cyclic, tridiagonal matrix, from which one can easily show that the eigenvalues all condense onto $\pm \epsilon$ for all other cases. A schematic of such a network, which we examine in the following subsections, is shown at the top left of Fig.~\ref{fig:8_infinity}. 

\subsection{From resolvent to eigenvalue distribution: an electrostatics connection}

There are well-known connections~\cite{Metz2019, shnerb1998winding, Derrida2000} between non-Hermitian random matrix theory and 2d electrostatics in the complex plane. For example, the trace of the resolvent (or Green's function) of a random matrix $\bf M$, 
\begin{equation}
\bf G(z) = \frac{1}{\bf{M} - z},
\end{equation}
is related to an electrostatics potential $\phi$, whose corresponding charge distribution gives us the eigenvalue density of states $\rho(z)$ in the complex plane:
{\small{
\begin{eqnarray}
\frac { 4 } { N } \partial _ { \bar z  } \operatorname { Tr } \mathbf { G }  ( z )& =&  - \frac{\partial^2}{\partial z\partial \bar z} \phi(x,y) \\
&=& - \left( \frac { \partial ^ { 2 } } { \partial x ^ { 2 } } + \frac { \partial ^ { 2 } } { \partial y ^ { 2 } } \right) \phi = - 4 \pi \rho(z),
\end{eqnarray}}}where $x$ and $y$ respectively denote the real coordinate and the imaginary coordinate, and $z$ and $\bar z$ denote a complex number $x + i y$ and its conjugate. Note that Tr$\bf G(z)$ is itself closely related to the \textit{electric field} associated with the charge distribution $\rho(z)$. 
For sparse, oriented, and locally tree-like graphs with no cycles, the trace of the resolvent can be calculated via the cavity method \cite{Metz2019}:
\begin{eqnarray} \label{eq:tree_trace}
\operatorname { Tr } \mathbf { G }  ( z )  = \sum_{j = 1}^N \frac { 1 } { d_ { j } - z }.
\end{eqnarray}
Note that this result depends on the random diagonal elements of $\bf M$, and is independent of any random off-diagonal elements of $\bf M$. According to Gauss's law of electrostatics, the charge distribution on an equipotential surface generates, in the region of space that is otherwise without charge, the same electric field that results when all charges act as if they are concentrated at the  origin of the complex plane. 

For a bidiagonal ``one-way'' matrix $\bf M$ (accessible by taking an appropriate $g \rightarrow \infty$ limit in Eq.~(\ref{eq:M}), see below) with no corner element (i.e. Eq.~(\ref{eq:M}) without periodic boundary conditions) and hence no cycles, Eq.~(\ref{eq:tree_trace}) tells us that $\rho ( z ) = 0 \text { for } | z | > \max _ { j } d _ { j }$~\cite{Metz2019}. This conclusion also holds for any non-cyclic bidiagonal matrix, where the eigenvalues simply take on the values of the diagonal elements $\lambda_j = d_j$. 

However, upon imposing periodic boundary conditions, the spectral distribution changes dramatically and leads to a rich variety of eigenvalue correlations and eigenvector localization within the spectrum. In the next subsection, we examine the model in Eq.~(\ref{eq:M}) with self-interactions $\epsilon \geq 0$ and strong (counter-clockwise) directional bias $g \rightarrow +\infty$. To have a well-defined limit, we rescale the matrix $\bf M$ in Eq.~(\ref{eq:M}), and study the properties of ${\bf{M'}} = e^{-g} {\bf{M}}$ in the limit $g \rightarrow \infty$, setting $\epsilon' = \epsilon e^{-g}$. Thus, we shall be interested in the spectra and eigenvalues of 
\begin{equation} \label{eq:Mprime}
{\bf{M'}} = \sum_{j} \left[ s _ { j } ^ { + } | j + 1 \rangle\langle j |  + \epsilon' d_j | j \rangle \langle j | \right],
\end{equation}
where $\epsilon'$ is fixed and $\left \{ s_j^+ \right \}$ and  $\left \{ d_j\right \}$ are random numbers drawn from the bimodal probability distributions $P_u(s)$ and $P_v(d)$ displayed in Fig.~\ref{fig:1_schm}. 

\subsection{Spectral curve confinement: another electrostatics connection}

\begin{figure}[h!]
\includegraphics[width=1\columnwidth]{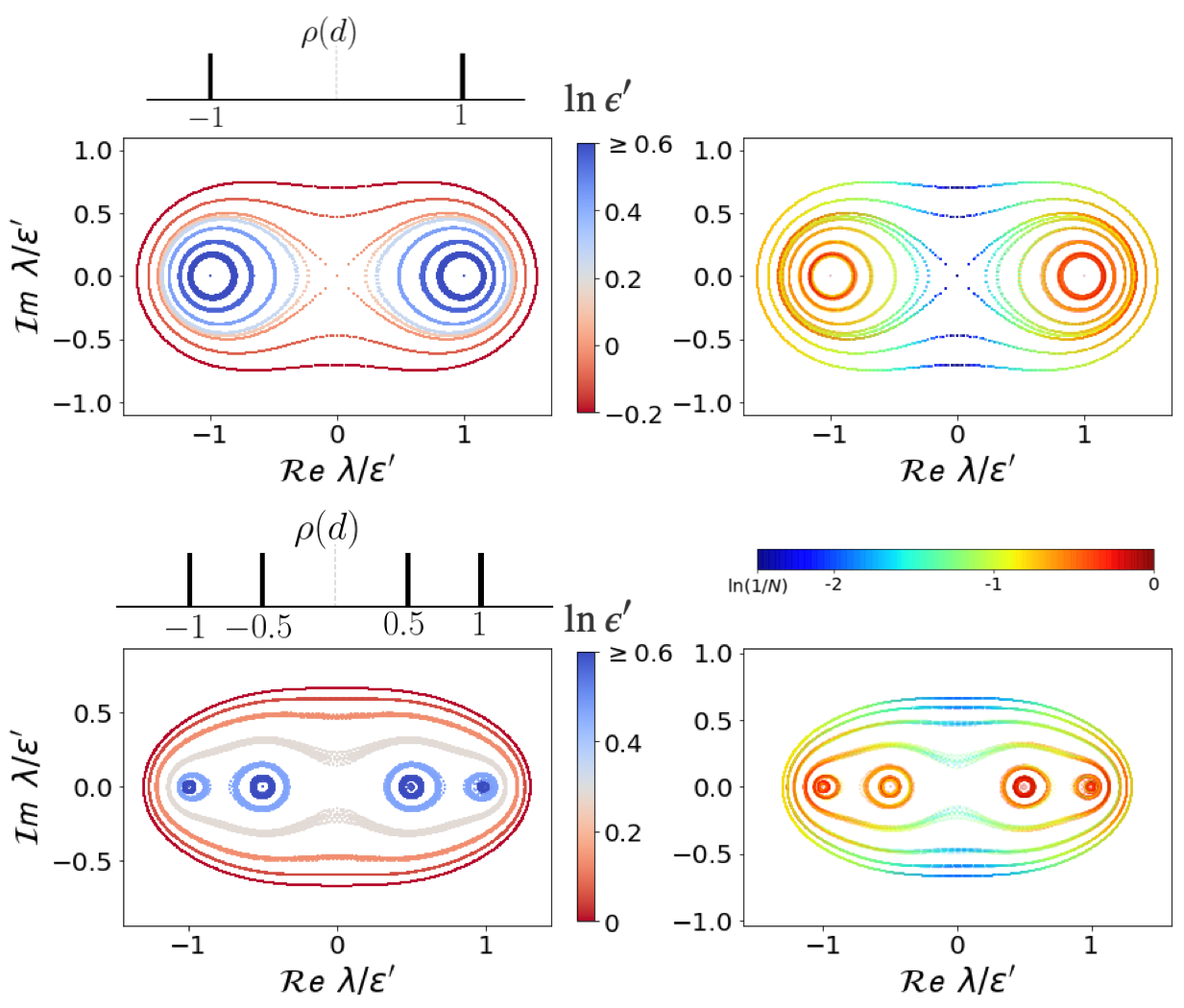}
\caption{\label{fig:7_orient} \linespread{1}\selectfont{ Left: When the network in Eq.~(\ref{eq:M}), rescaled as $\bf M'$ in Eq.~\ref{eq:Mprime}, has absolute directionality (the limit $ g \rightarrow + \infty$ in Eq.~(\ref{eq:M}) up to rescalings), the eigenvalue spectrum collapses onto a 1d curve that corresponds exactly to an equipotential surface resulting from a set of charges placed on the complex plane according to the probability distribution of the diagonal coefficients $P_v(d)$. The potential $V$ of the spectral curve is determined by $\ln \epsilon'$ (Eq.~(\ref{eq:charge_analogy})). Equipotential curves with negative potentials $V = \ln \epsilon' < 0$ expand outwards away from the two central charges and, in the limit of $\epsilon' \ll 1$, recovers the radially symmetric eigenvalue distribution for a zero-diagonal one-way hopping matrix. Top left: The diagonal coefficients follow the random sign distribution. In the limit of large $N$, the spectra correspond to the equipotential surfaces resulting from two like charges placed at $\pm 1$ on the real line. Bottom left: The distribution of diagonal coefficients corresponds to placing four like charges at $\pm 1$ and $\pm 0.5$ on the real line. Right panels: Inverse participation ratios, as random self-interactions start to dominate over hopping disorder ($\epsilon' \gg 1$), the spectrum becomes more localized with the most localized states occuring near the central charges; each equpotential curve is colored according to eigenvector IPR averaged over 20 realizations of matrices with rank $N=300$.}}
\end{figure}

As shown in Fig.~\ref{fig:7_orient}, upon imposing absolute directionality onto the network in Eq.~(\ref{eq:M}), the spectrum becomes confined to a one-dimensional locus in the 2d complex plane. As we now show, this curve is exactly the equipotential surface resulting from a charge distribution placed in the complex plane according to the \textit{probability} distribution of the diagonal elements $P_u(d)$, with the potential determined by the log-mean of ratio of the superdiagonal hopping magnitude $s^+$ to the strength of the rescaled diagonal disorder $\epsilon'$, $\langle \ln | s^+/\epsilon' |\rangle$. 

To derive this result, consider the oriented random connectivity matrix $\bf M'$ defined by Eq.~(\ref{eq:Mprime}) as displayed below,
\begin{equation}
{\bf{M'}} = 
\left[ \begin{matrix}
 \epsilon' d_1                          & s_1^+    &                        &         & &                   &  0 \\ 
0   &  \epsilon' d_2                         &   s_2^+ &           & &  0                &  \\
                           &                     & \ddots                 &   \ddots         &                 &   \\
           &               & 0  &   \epsilon' d_j                      &  s_{j+1}^+ &   \\
                           &    &               &        & \ddots    &   \ddots            &  \\
                           & 0   &                       & & 0 &   \epsilon' d_{N-1}                      &   s_{N-1}^+ \\
s_{N}^+   &    &                       &  &                      & 0 & \epsilon' d_{N} \end{matrix} \right],
 \end{equation}
and use a cofactor expansion to calculate the characteristic polynomial for the eigenvalues,  
 \begin{eqnarray} \label{eq:cofactor}
 \prod_{j = 1}^N \left( d_{j} - \frac{ \lambda}{\epsilon'} \right) &=& (-1)^{N-1} \prod_{j=1}^N  \left ( \frac{s_j^+ }{\epsilon'} \right ).
 \end{eqnarray}
We first multiply Eq.~(\ref{eq:cofactor}) by its complex conjugate, and then take the square root and logarithm of both sides. We then note that
\begin{eqnarray} \label{eq:spec_discrete}
\lim_{N\rightarrow \infty} \frac{1}{N} \sum_{j=1}^N \ln \left | \frac{s^+}{\epsilon'}  \right|  \equiv  \left \langle \ln \left | \frac{s^+}{\epsilon'} \right| \right \rangle,
\end{eqnarray}
where the RHS results from application of the law of large numbers. 
In the continuous limit, relabeling $ d_j \rightarrow d$, and rescaling $\dfrac{\lambda}{\epsilon'} \rightarrow \lambda$, we find that the spectral curves of Fig.~\ref{fig:7_orient} satisfy 
\begin{eqnarray} \label{eq:charge_analogy}
\int dd' \rho(d') \ln \frac{1}{|d' - \lambda|}   = -\left \langle \ln \left |\frac{s^+ }{\epsilon'} \right|  \right \rangle \equiv V,
\end{eqnarray}
where $\rho(d') \equiv P_v(d') $ is the probability distribution of the diagonal random variable. 

From Eq.~(\ref{eq:charge_analogy}), an analogy with two-dimensional electrostatics is immediately apparent: $\rho(d)$ is the distribution of like charges in the complex plane, while the potential $V$ experienced by a test charge on the equipotential surface is given by the $\log$-mean of the absolute value of the hopping variable times the ratio of the hopping interaction strength and the self-interaction strength:  $-\left \langle \ln \left | \frac{s^+}{\epsilon'}  \right| \right \rangle$. In the special case of the bimodal box distribution $P_u(s)$ with box distributions centered at $\pm1$ for the hopping matrix elements, $\langle \ln |s^+| \rangle = 0$, and the potential determining the spectral curve is just $\ln  \epsilon'$.  

This connection is explicitly illustrated in Fig.~\ref{fig:7_orient}. For a random sign diagonal distribution $P_{v=0} (d)$ and bimodal hopping term distribution $P_{u}(s)$ with equal self-interactions and hopping strengths $\epsilon' = 1$, $\rho(d) \equiv P_{v=0}(d) = \frac{1}{2} \left[\delta(d-1) + \delta(d+1)\right]$ and $P_u(s^+) = \mathcal{U}(-1-u, -1+u) + \mathcal{U}(1-u, 1+u)$, where $\mathcal{U}$ denotes the bounded uniform distribution shown in Fig.~\ref{fig:3_ginibre}. Eq.~(\ref{eq:charge_analogy}) for the eigenvalue distribution then assumes a particularly simple form,
\begin{eqnarray} \label{eq:equipotentialsurface_zero}
\frac{1}{2} \left( \ln |1 - \lambda| + \ln |1 + \lambda| \right )&=& 0,
\end{eqnarray}
which explains the infinity-shaped spectral shapes shown in the top panels of Fig.~\ref{fig:7_orient}. This relation holds true regardless of the value of the box-width $u$ of bimodal distribution of the nearest neighbor connections $s^+$, as long as $N$ is large enough for central limit theorem to apply. 
Eq.~(\ref{eq:equipotentialsurface_zero}) reveals that the complex eigenvalues must lie on the $V=0$ equipotential surface resulting from two like charges placed at $\pm 1$ on the real line.

On the other hand, if the random self-interaction strength becomes stronger than the hopping strength $\epsilon' > 1$, then the potential $V$ of the equipotential curve as indicated on the RHS of Eq.~(\ref{eq:charge_analogy}) becomes positive. With increasing $\epsilon'$, the eigenvalues condense onto equipotential surfaces ever-closer to the central source charges on the real line. 

When the nearest-neighbor connections instead exceed the self-interaction strength, $\epsilon' < 1$, the potential $V$ of the equipotential curve decreases to negative values and the spectral curves expand farther away from the charges determined by the diagonal elements. In the limit of $\epsilon' \ll 1$, i.e. $V \rightarrow -\infty$, the complex eigenvalues are large enough so that the charge distribution created by the diagonal disorder appears as a single point charge at the origin, which recovers the radially symmetric eigenvalue distribution for a zero-diagonal one-way hopping matrix \cite{Amir2016}. 

Eq.~(\ref{eq:charge_analogy}) holds true for any probability distribution $\rho(d)$ of the diagonal element. The bottom panels of Fig.~\ref{fig:7_orient} show the evolution of equipotential spectral curves for variable $P_u(s)$ and with the diagonal probability distribution $\rho(d) = \frac{1}{4} \left[ \delta(d-1)+ \delta(d-\frac{1}{2}) + \delta(d+\frac{1}{2}) + \delta(d+1) \right]$.

It is important to note, however, that although Eq.~(\ref{eq:charge_analogy}) tells us where the eigenvalues are \textit{allowed} to be---on an equipotential curve--it does not reveal how they are distributed on the curve, nor does it reveal how the IPR behaves on this curve. These issues are addressed in the next section.

\subsection{Continuous evolution of eigenvector localization length and eigenvalue correlations along the spectral curve}

\begin{figure}[h!]
\includegraphics[width=1\columnwidth]{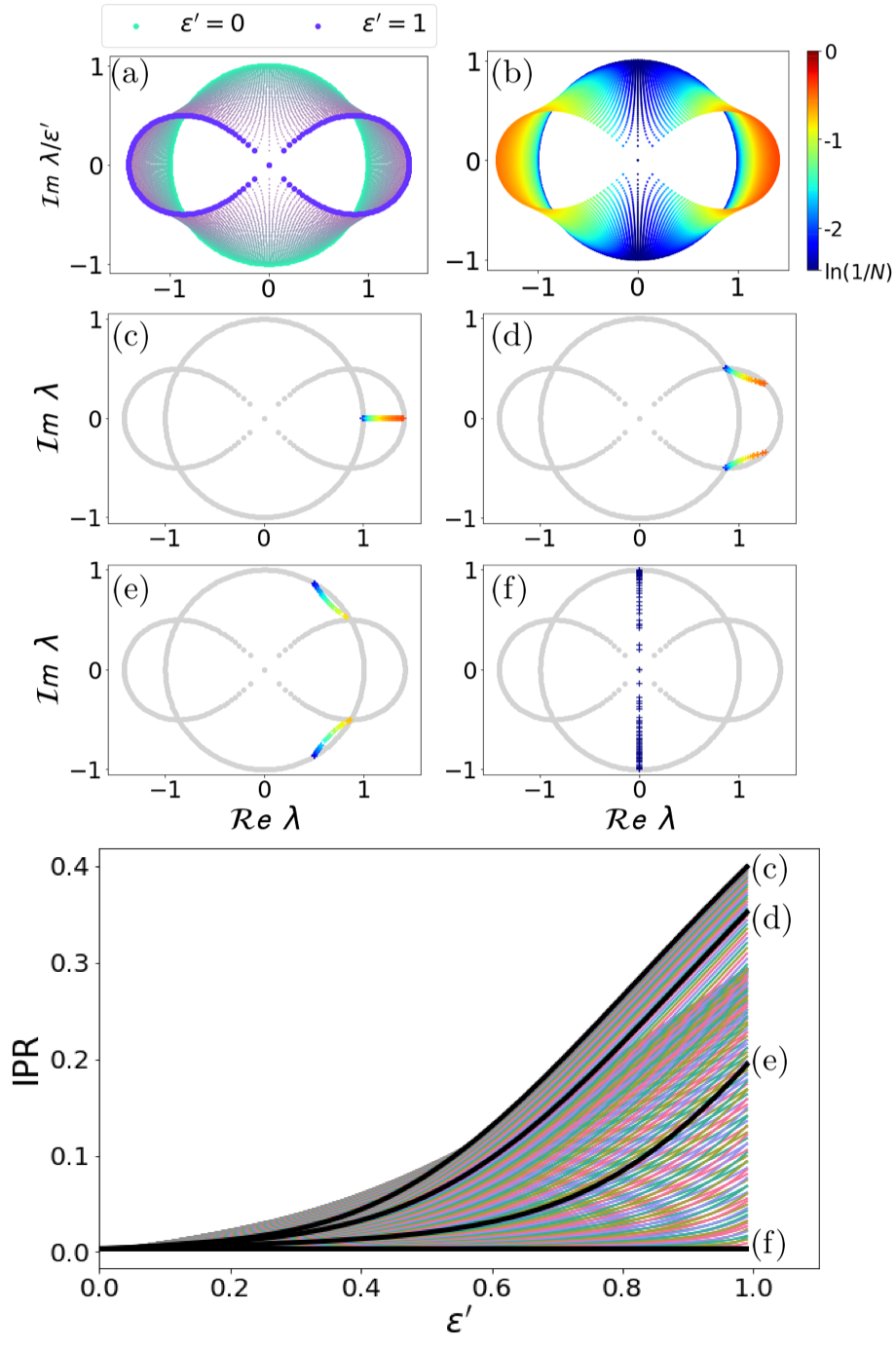}
\caption{\label{fig:9_wavenumber} \linespread{1}\selectfont{(a) and (b): The evolution of eigenvalue positions (left) and right eigenvector localization measured by the IPR$_R$ (right) are shown for a fixed set of $\pm 1$ self-interacting and $\pm1$ hopping elements ($N=300$ matrices drawn from $P_{v=0}(d)$ and $P_{u=0}(s)$), but with the relative strength of the self-interactions $\epsilon'$ tuned from 0 (green points) to 1 (purple points). (c)--(f): The eigenfunction localization behavior, quantified by $\ln \text{IPR}_R(\lambda_{k_n})$, as a function of $\epsilon'$, following seven individual eigenvalue trajectories originating from four distinct wave numbers $k_n$ in the $\epsilon' = 0$ spectrum. These images suggest that the $\epsilon' = 0$ wavenumbers play a role in ``assigning'' localization properties to the eigenvectors of certain eigenvalues as the nonzero self-interactions are turned on. The eigenvector of the eigenvalue with the largest real part localizes drastically. On the other hand, The eigenvector of the eigenvalue closest to 0 stays delocalized. The family of curves at the bottom show the evolution of  $\ln \text{IPR}_R(\lambda_{k_n})$ for all $k_n$ values as $\epsilon'$ increases from 0 to 1.}}
\end{figure}

\begin{figure}[h!]
\includegraphics[width=1\columnwidth]{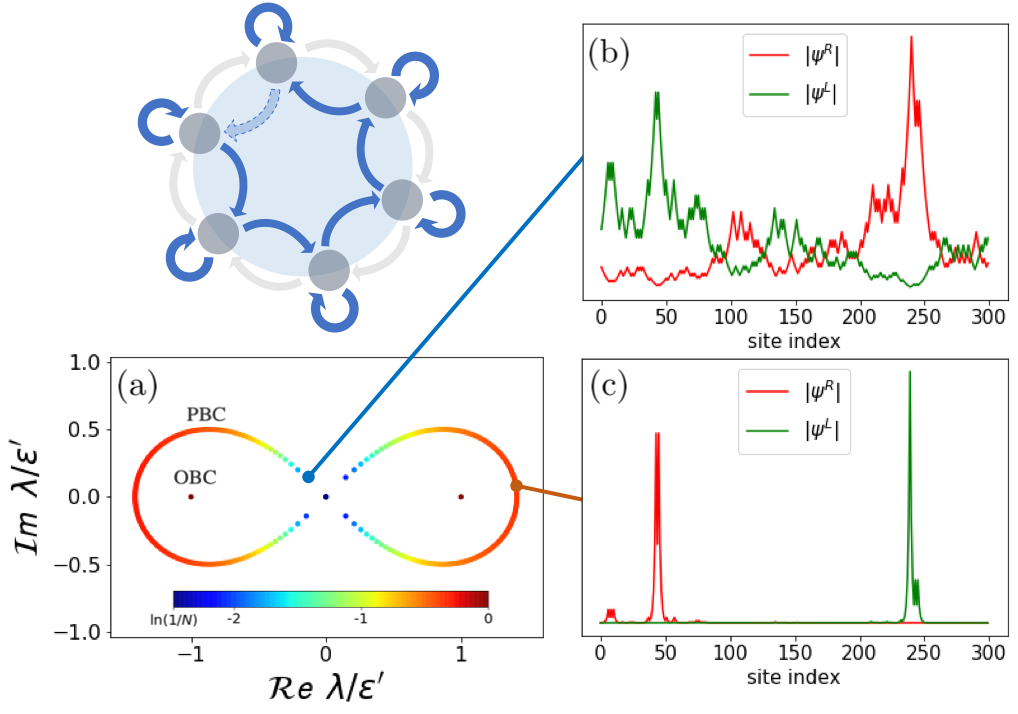}
\caption{\label{fig:8_infinity} \linespread{1}\selectfont{(a): Comparison of the spectra under open boundary conditions (OBC) and periodic boundary conditions (PBC) corresponding to one realization ($N=300$) of the matrix $\bf M'$ shown in Eq.~(\ref{eq:Mprime}), which arose from the large $g$ limit of Eq.~(\ref{eq:M}), with equal strength random hopping and random self-interactions ($\epsilon' = 1$). Breaking a single link of the ring (shown schematically as the dashed arrows in the top left) changes the spectrum entirely; the spectrum condenses onto two points with degeneracy $N/2$ when the cycle is broken. Eigenfunction amplitudes of right and left eigenvectors $|\psi^R |$ and $|\psi^L |$ corresponding to (b) an eigenvalue close to 0 and (c) the eigenvalue with the largest real part in the spectrum corresponding to periodic boundary condition (PBC). The eigenvectors in (b) and (c) show a drastic difference in their degrees of localization. In addition, the left and right eigenvectors corresponding to the same eigenvalue show different centers of localization.}}
\end{figure}

It is known that the eigenspectra for the directional networks, with $s_j^+ = \pm 1$ random hopping but no disorder on the diagonal, have strongly delocalized eigenvectors similar to plane waves~\cite{Amir2016}. Indeed, after making a simple similarity transformation determined by the paricular realization of the superdiagonal disorder, eigenvalues and eigenvectors can be found analytically to be
\begin{eqnarray} 
{ \lambda _ { k _ { n } } = e ^ { g + i k _ { n } } } \label{eq:lambda_pert}\\ 
{ \langle x | \psi _ { k _ { n }  } \rangle = \frac { 1 } { \sqrt { N } } e ^ { i k _ { n } x } } \label{eq:evec_pert}
\end{eqnarray}
where
\begin{eqnarray} \label{eq:wavenumber}
 k _ { n } = \frac { 2 \pi n } { N } , \quad n = 0,1,2 , \cdots , N - 1. 
\end{eqnarray}
Note that Eq.~(\ref{eq:lambda_pert}) and (\ref{eq:evec_pert}) imply extended eigenfunctions and a spectrum with an elliptical shape in the complex plane in the limit $\epsilon' \rightarrow 0$. 

Remarkably, however, upon incorporating nonzero self-interactions with strength $\epsilon' = 1$ with a random sign distribution $P_{v=0}(d)$, we find that the spectra not only transforms into the distinct shape of an infinity symbol (as shown in Fig.~\ref{fig:7_orient}), but also acquire right eigenstates with an entire range of localization lengths (see the two eigenfunctions shown in Fig.~\ref{fig:8_infinity}). The top of Fig.~\ref{fig:9_wavenumber} shows, for a fixed set of $\pm 1$ self-interactions and $\pm 1$ hopping elements drawn from $P_{v=0}(d)$ and $P_{u=0}(s)$ respectively, the variation in eigenvalue position (left) and eigenvector localization (right) as the self interaction strength $\epsilon'$ is tuned from 0 to 1. The rest of Fig.~\ref{fig:9_wavenumber} shows how the localization indicator $\ln |\text{IPR}_R(\lambda_{k_n})|$ evolves as a function of $\epsilon'$ for seven individual trajectories originating from four distinct wave numbers $k_n$, shown in Eq.~(\ref{eq:wavenumber}) for the $\epsilon' = 0$ spectrum. Similar to the winding numbers studied in Ref.~\cite{shnerb1998winding}, these wavenumbers can be used to classify localization properties of the eigenvectors in the presence of nonzero self-interactions. Importantly, when self-interactions are incorporated into the oriented ring network, the principal eigenvector (the eigenvector corresponding to the eigenvalue with the largest real part), transforms from being completely delocalized to being highly localized (see Fig.~\ref{fig:8_infinity}c and Fig.~\ref{fig:9_wavenumber}). Such eigenvectors dominate long term dynamics of systems linearized about some equilibrium state (see next section and Appendix \ref{app:3_leftright}). As shown for the ring network studied in Ref.~\cite{Tanaka2018}, the principal eigenvector of the connectivity matrix of a neural network dictates the sustained activity associated with short term memory, and the presence of a localized principal eigenvector is associated with short term memory storage of information regarding a specific spatial direction. 

We leave for future work an elucidation of both the density of states along this family of continued spectral curves and the intriguing continuous variation of eigenvector localization properties. We emphasize again that both strong directionality and cycle lengths of the order of the system size appear to be necessary ingredients for producing these striking spectra. 

\subsection{Consequences of nontrivial distinction between left and right eigenvectors} \label{sec:leftright}
In this work, we have focused on the localization properties of the right eigenvectors $\psi_n^R(j)$, which appear naturally in, say, neural dynamics problems such as Eq.~(\ref{eq:neural_linear}) when the firing rate is expanded according to $r_j(t) = \sum_n c_n \psi_n^R(j) e^{\lambda_n t}$. In systems with no directional bias (e.g. the random sign model examined in Sec.~\ref{sec:repulsion}), we find that the left and right eigenvectors are identical (see Appendix \ref{app:3_leftright}). However, when nonzero directionality bias is present, left and right eigenfunctions corresponding to the same eigenvalue can differ in both localization lengths and the positions of their centers of localization (see Fig.~\ref{fig:8_infinity}). This dichotomy has interesting consequences, both formally and physically. Formally, the distinction between left and right eigenvectors means that one can define an alternative metric of localization using left-right eigenvector inner products, 
\begin{eqnarray}
IPR_{LR}(\lambda_n) = \left[\frac{ \left( \sum_i \left|\psi^L_n(i) \psi^R_n(i) \right| \right)^2}{\sum_i \left|\psi^L_n(i) \psi^R_n(i) \right|^2 } \right]^{-1},
\end{eqnarray}
which yields different results from the IPR defined by Eq.~(\ref{eq:IPR}) for systems with nonzero directional bias. 
Physically, the spatial separation between left and right eigenvectors manifests in a nontrivial distance between the response and excitation signals in a neural network. This can be seen via the signal propagator. For zero input $h_i(t)$ in Eq.~(\ref{eq:neural_linear}), the propagator takes the following form~\cite{hatano1998non}
\begin{eqnarray}\label{eq:prop}
G_{ij}(t) = \sum_{n = 0}^{N-1} \psi_n^R(i) \psi_n^L(j) e^{\lambda_n t - t/\tau}.
\end{eqnarray}
Upon applying the biorthogonality property of left and right eigenvectors with proper normalization $\sum_{n} \psi_n^R(i) \psi_n^L(j) = \delta_{i,j}$, one can verify that Eq.~\ref{eq:prop} reduces to the Kronecker delta function $\delta_{i,j}$ at $t=0$. As seen in Eq.~(\ref{eq:prop}), for a pair of right and left eigenvectors peaked respectively at $i$ and $j$, an excitation signal at $j$ triggers a response at $i$. When directional bias is nonzero, the right and left eigenvectors peak at different locations separated in space. Thus, even when the left and right eigenvectors are individually localized, they can communicate over a large spatial region. 
Various aspects of the left and right eigenvectors are further explored in Appendix~\ref{app:3_leftright}.

\section{Discussion and outlook} \label{sec:conclusion}
\subsection{Implications for more complicated networks }
Knowledge of the spectral properties of sparse non-Hermitian random matrices is critical for determining the behavior of real-world networks and also necessary for devising practical methods for understanding network data. In this paper, we briefly highlighted the implications of our results for the dynamics of a ring neural network. However, spectral properties of eigenvalues and eigenvectors can be key for analyzing other types of networks, including those with a percolation threshold and networks where the relative importance (centrality) of nodes plays a key role~\cite{newman2010networks}. 

Focusing on a simple one-dimensional network allowed us to identify important network ingredients, such as random nearest neighbor connections, random self-interactions, large loop structure, and strong directional bias.  
Ideas and results from this paper may shed light, on different levels, on more general networks with higher node degrees (dimensionality) and structure types, as summarized below.

\subsubsection{Eigenvalue repulsion and eigenvector localization}\label{sec:nonherm}

The phenomena of eigenvalue repulsion arising only for more extended eigenstates (Sec.~\ref{sec:repulsion_localization}) might conceivably be a property of all non-Hermitian random matrices, regardless of their degree of sparsity and underlying spatial structure. Although the numerical evidence in this paper is consistent with this conjecture, new mathematical tools may be necessary to prove this connection convincingly. Note that although the inverse participation ratio (IPR) may no longer be precisely the inverse of a physical localization length outside of strictly one-dimensional networks, it can still be used as a measure of the inverse cluster size for eigenmodes in more general sparse networks. 

\subsubsection{Properties of random 1d systems}
In Sec.~\ref{sec:repulsion_localization} and Sec.~\ref{sec:self}, we studied how nonzero random self-interactions and a spreading distribution in nearest neighbor connection strengths lead to eigenvector localization and eigenvalue decorrelation for one-dimensional models. When randomness in the self-interactions is strong enough, all eigenvectors become highly localized and there are negligible correlations among the eigenvalues in the spectra, though strong directional bias (large $g$ in Eq.~(\ref{eq:M})) makes the system more resistant to this outcome. 

Unlike the more general conjectures in the section immediately above, our computations focused on a one-dimensional network represented by a tridiagonal random matrix, with corner matrix elements used to implement periodic boundary conditions. Random graphs with this structure can arise in the analysis of certain complex systems, which contains spatial scales where spatially-local couplings are prevalent. Examples include biological networks such as the ring attractor neural network~\cite{Kim2017}, matrices that describe DNA single-nucleotide polymorphism data~\cite{paschou2007pca}, and even complex systems in economics, which can sometimes be approximated by decomposable matrices~\cite{SimonAndo1961,MeyerDataClustering,Cucuring2013EigenvectorMigration}. An alternative set of unidimensional systems also arise naturally in the temporal ordering of time series data~\cite{Cucuringu2011}. In these areas, eigenvector localization properties are critical to the functioning of different spectral algorithms used for detecting network boundaries and temporal patterns~\cite{Martin2014EigenvectorLocalization,Cucuring2013EigenvectorMigration,Mavroeidis2012EigenvectorPCA,Taylor2017EigenvectorTemporal}.

\subsubsection{Extendibility to low-dimensional graphs}

It would be interesting to explore whether the results of the one-dimensional models studied in this paper are extendible to similarly-structured networks with more degrees of freedom associated with each node, similar to the 3 sites per node model studied in Ref.~\cite{Amir2016}. More generally, we can ask: What is the effect of competing self-interaction disorder and connectivity disorder on eigenvector localization and eigenvalue repulsion in low-dimensional graphs?

Low-dimensional graphs are ubiquitous in nature and appear often in the form of planar networks, such as leaf vasculature and water networks \cite{katifori2012}. Both diffusion between nodes (analogous to the hopping terms in Eq.~(\ref{eq:M})) and directed motion (controlled by the parameter $g$ for Eq.~(\ref{eq:M})) appear naturally in these models. The higher-dimensional connectivity embodied in the branching networks, however, raises interesting questions, such as different ways of distributing the connection number per node within a connected space.

\subsection{Challenges in mathematics}
Rigorous results for sparse non-Hermitian matrices are difficult to obtain since it is challenging to apply standard random matrix theory tools, and proofs of convergence for eigenvalues and eigenvectors in the thermodynamic limit of large rank matrices remain elusive~\cite{Metz2019}. 

In this work, we explored properties of sparse non-Hermitian random matrices, predominantly through numerical random matrix experiments that highlight the need for more precise mathematical descriptions. 

One question is: How can we derive a mathematical description of the relation between complex eigenvalue repulsion and eigenstate delocalization? Eigenvalue repulsion and eigenstate delocalization with interactions in sparse Hermitian systems are currently of interest in quantum many-body systems \cite{Nandkishore2015,pal2010}; it would be interesting to see if methods developed for quantum systems could be carried over to non-Hermitian systems. 

Another question is, what is the effect of large cycles on the spectral properties of sparse random matrices? An important technique used for calculating the spectral density of sparse non-Hermitian random matrices is the cavity method \cite{Rogers2009}, which has been successful in determining the spectral gap and distribution of outlier eigenvalues and eigenvectors. The presence of large cycles, however, breaks the method's assumption of a local tree-like structure. Another potential analytical relation is the non-Hermitian generalization of the Thouless relation relating the localization length and the density of states~\cite{Thouless_1972}, which also exploits an electrostatics analogy (in terms of the Lyapunov exponent) for random one-dimensional systems \cite{Derrida2000}. However, it may be challenging to apply this method in the combined presence of periodic boundary conditions and delocalized eigenfunctions. Many works have examined the effects of small cycles (cycles with constant number of nodes that do not grow with system size) \cite{Metz2011, bolle2013spectra,coolen2016replica,newman2019}. Nevertheless, as shown in this paper, the presence of a single large cycle on a sparse graph can drastically change the system's spectral correlation and localization behavior, making large cycles a worthwhile problem for future studies.

\begin{acknowledgments}
It is a pleasure to acknowledge helpful conversations with A. Amir, J. Kates-Harbeck, and B. Shklovskii. We also thank A. Amir for a critical reading of the manuscript. G.H.Z. acknowledges support by the National Science Foundation Graduate Research Fellowship under Grant No. DGE1745303. This work was also supported by the National Science Foundation, primarily through Grant No. DMR-1608501 and through the Harvard Materials Science and Engineering Center, via Grant No. DMR-1420570.
\end{acknowledgments}

\appendix

\section{Pair correlation function $g(r)$ calculation for eigenfluids \label{app:1_gr}}

In this section, we describe in detail our numerical extraction of pair correlation functions. We first review the pair correlation function for a homogeneous fluid, appropriate for the Ginibre random matrix ``eigenliquid'', and then generalize the procedure for inhomogeneous eigenvalue distributions. 

\subsection{Pair correlation function of the homogeneous eigenliquid generated by the Ginibre ensemble}

For an isotropic homogeneous fluid, the particle density of one realization is given by
\begin{eqnarray}
\rho(r) = \rho g(r),
\end{eqnarray}
where $\rho$ is the average density of eigenvalues in the complex plane, and $g(r)$ is the probability of finding a particle distance $r$ away from a reference particle at the origin in one realization of the ensemble. Upon integrating $\rho(r)$ over a ring of small width $dr$ at distance $r$ away from the central particle, we have
\begin{eqnarray}
n(r) dr &\approx& \rho g(r) 2 \pi r dr, 
\end{eqnarray}
where $n(r) dr$ is the number of particles between $r$ and $r+ dr$ about the central particle. For $N$ total particles in the realization, the number of particle pairs that are separated by distances between $r$ and $r + dr$, which we denote $G(r, dr)$, is then given by 
\begin{eqnarray}
G(r, dr) = \frac{N}{2} n(r) dr = \frac{N}{2} \rho g(r) 2 \pi r dr. 
\end{eqnarray}

Thus, the pair correlation function (or radial distribution function) $g(r)$ is given by~\cite{mcquarrie2000statistical}, 
\begin{eqnarray} \label{gr_norm_homo}
g(r) = \frac{G(r, dr)}{N \rho \pi r dr}, 
\end{eqnarray} 
where we found $G(r, dr)$ numerically using a binary search tree ($\tt{CKD\_Tree}$ python package). 
We tested this procedure for the Ginibre ensemble, for which the scaled eigenvalues are contained in a disk of radius 1, and the eigenvalue density is $\rho = N/\pi$ everywhere inside this unit disk. Then, Eq.~\ref{gr_norm_homo} leads to a radial distribution function for the Ginibre ensemble $g_G(r)$ given by
\begin{eqnarray}
g_{\text{G}}(r) = \frac{G_{\text{G}}(r, dr)}{N^2 r dr},
\end{eqnarray} 
which we used to obtain the pair correlation function shown in Fig.~\ref{fig:3_ginibre}.

\subsection{Pair correlation function of inhomogeneous eigenfluids}

For an inhomogeneous, anisotropic fluid in two dimensions, the particle density of a single realization takes a more general form
\begin{eqnarray}
\rho(\vec r_2) = \rho(\vec r_1) g(\vec r_1, \vec r_2 - \vec r_1),  
\end{eqnarray}
where $g(\vec r_1, \vec r_2)$ is a probability distribution that depends on 4 coordinates, and $\rho(\vec r_2)$ is the chance of finding a particle at $\vec r_2$ given that there is a particle at $\vec r_1$ in the same realization. Upon performing a change of coordinates from $(\vec r_1, \vec r_2)$ to $(\vec r = \vec r_2 - \vec r_1, \vec R = (\vec r_1 + \vec r_2)/2)$, where $\vec r$ is the separation vector between the two particles and $\vec R$ is the mean location of particle pair, we consider eigenvalue correlations in a small area about $\vec R$, over which the eigenvalue density is appropriately constant. Following the same procedure as in the previous section, for sufficiently isotropic correlations where the angular dependence can be neglected (see Sec.~\ref{app:2_angular} below), we arrive at
\begin{eqnarray} \label{gr_norm_inhomo}
g(r, \vec R) =  \frac{G(r, dr, \vec R)}{N\rho(\vec R-\frac{\vec r}{2}) \pi r dr},
\end{eqnarray} 
where the local pair correlation function $g(r, \vec R)$ is the probability of finding two particles distance $r$ apart given that their mean location is $\vec R$, and $N$ is the number of eigenvalues in the reference area. 

To properly examine the local correlation function averaged over a small grid in space,  $g(r, \vec R)_{\Delta R}$, the box size $\Delta R$ should be much larger than the average particle spacing but small enough such that the fluid contained in the box is approximately homogeneous. In this case, we can apply Eq.~(\ref{gr_norm_homo}) and obtain,
\begin{eqnarray}\label{single_box_norm}
G(r, dr, \vec R)_{\Delta R} &\approx& \frac{N_{\vec R, \Delta R}^2}{\Delta R^2} ~ \pi r  dr 
\times g(\vec r, \vec R)_{\Delta R},
\end{eqnarray}
where $N_{R, \Delta R}$ is the number of particles in the box of size $\Delta R$ centered at $\vec R$, and $G(r, dr, \vec R)_{\Delta R}$ is the total number of particle pairs separated by distances between $r$ and $r+dr$, averaged over mean locations inside the reference box. 

Partitioning of the eigenfluid into smaller, approximately homogeneous boxes requires averaging over many realizations of the ensemble (diagonalization of many matrices) in order to achieve an adequate amount of statistics. Then, obtaining the proper normalization via comparison to an uncorrelated eigenfluid,
\begin{eqnarray} \label{eq:normg}
G_{0}(r, dr, \vec R)_{\Delta R} & = & \frac{1}{M} \frac{\sum_{j = 1}^M  (N^{(j)}_{R, \Delta R})^2}{2 \Delta R^2} ~2 \pi r  dr,
\end{eqnarray}
where $M$ is the total number of realizations, the local pair correlation function is 
\begin{eqnarray} \label{eq:localgrfinal}
g(r, \vec R)_{\Delta R} \equiv \frac{G(r, dr, \vec R)_{\Delta R}}{G_{0}(r, dr, \vec R)_{\Delta R}}. 
\end{eqnarray}
To calculate $g(r, \vec R)_{\Delta R}$, the probability of finding particles with separation distances $\in [r, r+dr)$ given that the mean locations of the particle pairs are within a box of size $\Delta R$ centered at $\vec R$, we thus find the numerator numerically via binary search trees, and calculate the denominator from Eq.~(\ref{eq:normg}). 

For this method to work, we require that $g(r, \vec R)_{\Delta R}$ does not change significantly with the box size $\Delta R$. If $\Delta R$ is too large, we average over areas with significantly different correlations (or eigenvalue densities), and the homogeneity assumption fails. We also avoid applying this method near the fractal edges of the random sign spectrum (see Fig.~\ref{fig:2_frac}), where the density of states change abruptly, and there are fine, singular density spikes in the spectrum. 

\subsection{Angular dependence of the pair correlation function} \label{app:2_angular}

In deriving Eq.~(\ref{gr_norm_inhomo}), we assumed that the angular dependence of the local pair correlation function $g(\vec r, \vec R) = g(r, \theta, \vec R) \approx g(r, \vec R)$ can be neglected, thus improving our statistics by counting all particle pairs separated by distance $r$ regardless of the direction of their separation vector. More generally, however, there could exist eigenfluids where correlations between particles can have significant dependence on the direction. Here, we explore angular dependence of the pair correlation function for the random sign model studied in Sec.~\ref{sec:repulsion}, and show that the angular dependence of the correlations is weak, justifying the approximation leading to Eq.~(\ref{gr_norm_inhomo}). 

\begin{figure}[h]
\includegraphics[width=0.95\columnwidth]{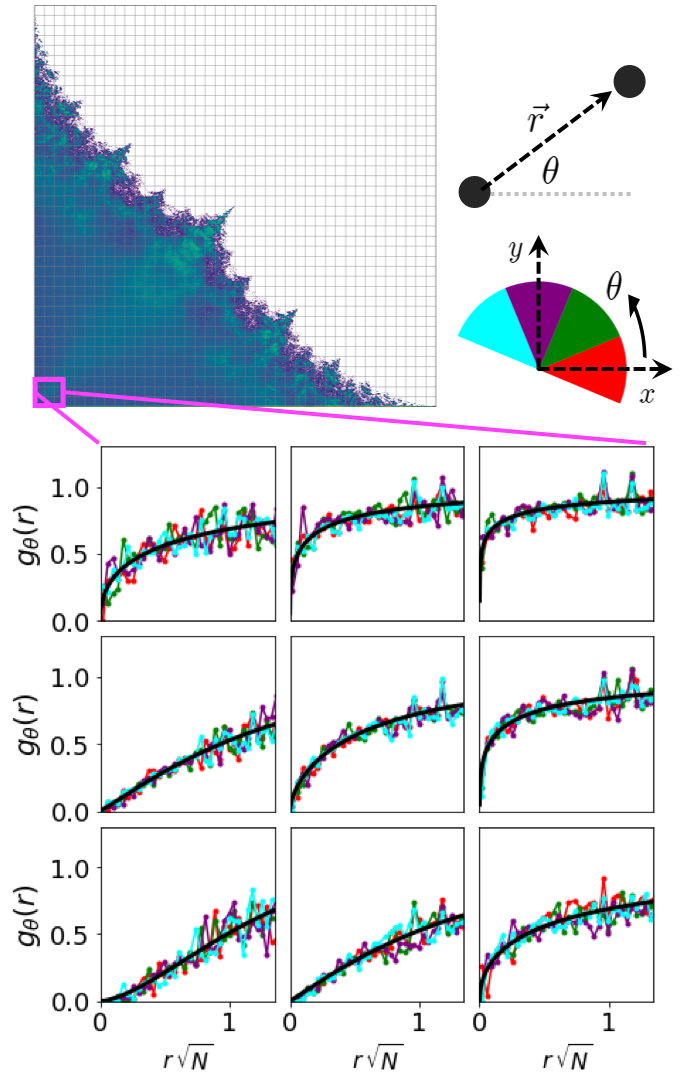}
\caption{\label{fig:A1} \linespread{1}\selectfont{Directional variation of the pair correlation function in the first quadrant of the random sign model, obtained from counting eigenvalue pairs centered in the 9 ($0.05 \times 0.05$) square grids closest to the origin, enclosed by the magenta box in the top left spectra. The top right schematic shows four different angular sectors (centered at 0, $\pi/4$, $\pi/2$, and $3 \pi/4$) within which the angle of an eigenvalue pair separation $\theta$ can be binned. In the bottom plots, the color of each line corresponds to the pair correlation function derived from counting eigenvalue pairs within that angular range. The smooth black line is the fit of the functional form $g_s(r)$ shown in Fig.~\ref{fig:3b_gr_fit}, i.e. angular-averaged pair correlation function. Since the colored lines do not deviate significant from the black line, isotropy of the eigenvalue correlations appears to be a reasonable approximation for the random sign model.}}
\end{figure}

To test for directional variation of the two-point eigenvalue correlations, we again examine correlations of eigenvalue pairs within the 9 ($0.05 \times 0.05$) square grids closest to the origin in the first quadrant, examined previously in Fig.~\ref{fig:3b_gr_fit}. The top left spectrum of Fig.~\ref{fig:A1} shows these regions enclosed by a magenta box. We bin eigenvalue pairs in these regions, based on the angle $\theta$ characterizing each of their separation vectors $\vec r$, into four angular sectors (see top right of Fig.~\ref{fig:A1}). Since eigenvalue correlations should be symmetric under $\theta \leftrightarrow -\theta$, we only study $\theta$ spanning a range of $\pi$. In the bottom plots, the color of each line then corresponds to the pair correlation function derived from counting eigenvalue pairs within that angular range, normalized as in Eq.~(\ref{eq:localgrfinal}) and multiplied by 4 (since we are binning into four angular sectors). The black smooth line is the fitting function $g_s(r)$ used in Fig.~\ref{fig:3b_gr_fit}, i.e. the rotationally averaged pair correlation function, obtained by counting all eigenvalue pairs of separation $r$ regardless of direction. Although the colored angular counts are noisier due to the reduction in sample size, we find no significant angular dependence. Finding and examining spectra for which the two-point eigenvalue correlations do exhibit nontrivial angular dependence would be an interesting topic for future investigations. 

\section{Eigenstate localization for nonidentical left and right eigenvectors} \label{app:3_leftright}

\subsection{Left and right eigenvectors with directional bias but no diagonal disorder}

\begin{figure}[h!]
\includegraphics[width=1\columnwidth]{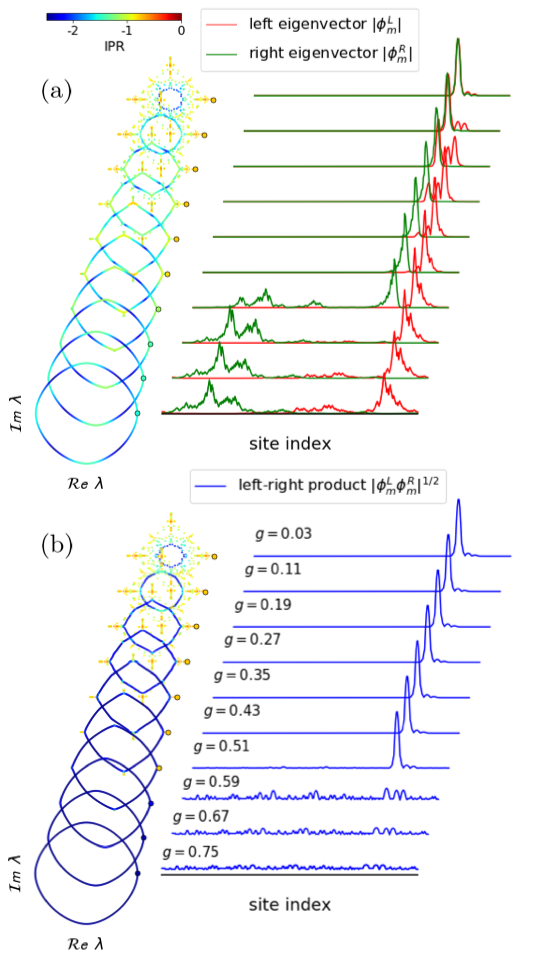}
\caption{\label{fig:A2_g} \linespread{1}\selectfont{Comparison of $IPR_R$ and $|\psi^R|$ and $|\psi^L|$ (top), and $IPR_{LR}$ and $|\psi^L \psi^R|^{1/2}$ (bottom) corresponding to the principal eigenvalue for one realization of the asymmetric random sign model with no self-interaction disorder, with increasing $g$. As $g$ increases from 0, the spectrum exhibits a band gap at the origin of the complex plane with a rim of weakly delocalized states, and the right and left eigenvectors spatially separate and gradually spread out. When $g$ is sufficiently large such that the spectral rim reaches the principal eigenvalue, the left and right eigenvectors experience a jump in separation and the left-right inner product $|\psi^R|$ and $|\psi^L|$ are abruptly delocalized.}}
\end{figure}

\begin{figure}[h!]
\includegraphics[width=1\columnwidth]{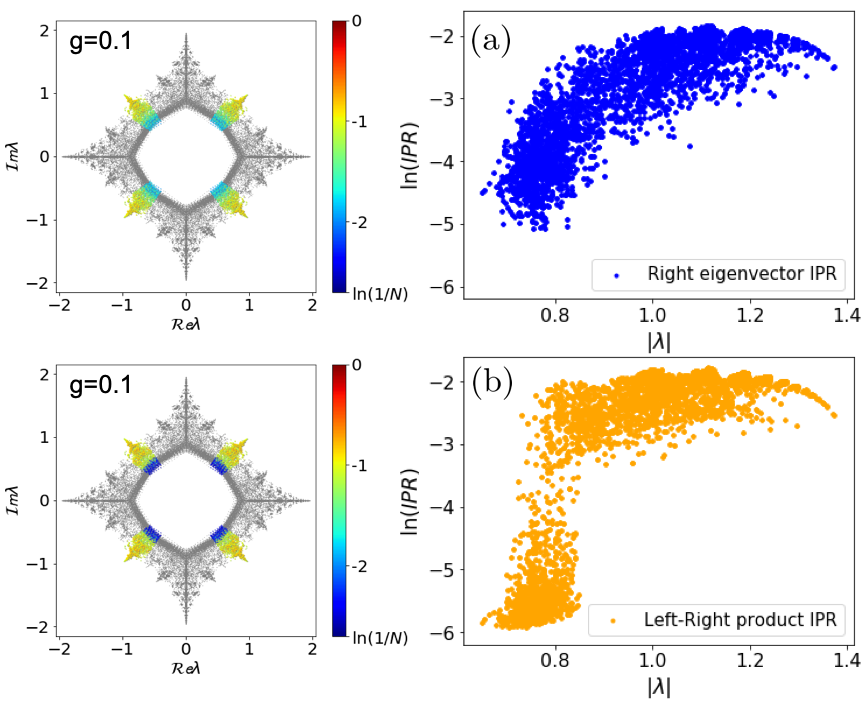}
\caption{\label{fig:A2b_IPR_jump} \linespread{1}\selectfont{IPR$_R$ (top) and IPR$_{LR}$ (bottom) as a function of the eigenvalue magnitude (distance away from the origin) for eigenvalues around the $45^\circ$ line in the complex plane, as indicated by the colored portions of the spectrum (left). Data extracted from 150 diagonalizations of the random sign hopping matrix $N=500$ (right) show that IPR$_R$ decreases gradually as the eigenvalue gets closer to the rim, while IPR$_{LR}$ abruptly drops when the eigenvalue is right at the rim.}}
\end{figure}

For a non-Hermitian matrix, the left and right eigenvectors are in general not identical. There then exist three distinct ways to measure localization for an eigenstate, by using the magnitude of the right eigenvector (Eq.~(\ref{eq:IPR})), by using the magnitude of the left eigenvector, 
\begin{eqnarray}\label{eq:IPR_l}
IPR_{L}(\lambda_n) = \left[\frac{ \left( \sum_i \left|\psi^L_n(i) \right|^2 \right)^2}{\sum_i \left|\psi^L_n(i) \right|^4 } \right]^{-1},
\end{eqnarray}
or by using the product of the left and right eigenvectors~\cite{hatano1998non},
\begin{eqnarray} \label{eq:IPR_lr}
IPR_{LR}(\lambda_n) = \left[\frac{ \left( \sum_i \left|\psi^L_i \psi^R_i \right| \right)^2}{\sum_i \left|\psi^L_i \psi^R_i \right|^2 } \right]^{-1},
\end{eqnarray}
where $i$ labels the $i$-th site of the left or right eigenvector. 

In the one-dimensional systems that we study using Eq.~(\ref{eq:M}), the left and right eigenvectors can be related through the equation,
\begin{eqnarray} \label{eq:lr_g_relation}
\psi^L (g) = \psi^R (-g) {\bf SS^T},
\end{eqnarray}
where  ${\bf S}$ is a diagonal matrix with elements $S_{jj} = \prod_i^{j-1} \sqrt{\frac{s_i^-}{s_i^+}}$, responsible for transforming $\bf{M}$ into a symmetric matrix~\cite{Amir2016}. When there is no directional bias in Eq.~(\ref{eq:M}) (i.e. g = 0), and the hopping probability distribution is narrow, the left and right eigenvectors are identical up to sign flips at each site,
\begin{eqnarray}
\psi^L_n(i) = \pm \psi^R_n(i),
\end{eqnarray}
where $i$ labels some site index and the $\pm1$ is determined by the matrix ${\bf S S^T}$, which in turn depends on the particular realization of the disorder. Thus, without directional bias, the left and right eigenvectors with the same eigenvalue have the same magnitude at each site, and hence share the same localization properties. In this case, all three defintions of IPR (Eq.~(\ref{eq:IPR}), (\ref{eq:IPR_l}), and (\ref{eq:IPR_lr})) give the same result for every eigenvalue in the spectrum. However, when there is nonzero directional bias ($g\neq 0$ in Eq.~(\ref{eq:M})), the left and right eigenvectors separate spatially, and the results change. As shown explicitly in Sec.~\ref{sec:leftright}, the left and right eigenvectors corresponding to the same eigenvalue can take on entirely different shapes. Fig.~\ref{fig:8_infinity} shows the spectra of the one way hopping (infinite directional bias) and random sign self-interaction model, where the right and left eigenvectors can be separated by a significant distance on the ring.
In these cases, IPR$_{LR}$ returns significantly different values compared to IPR$_R$ and IPR$_L$. 

Fig.~\ref{fig:A2_g} compares the magnitude of the left eigenvector and the right eigenvector, and the square root of the product of left and right eigenvectors for the asymmetric random sign model with no self-interaction disorder. The localization properties of this spectrum for just the right eigenvectors were studied in detail in Ref.~\cite{Amir2016}. We show a sequence of spectra with increasing $g$, focusing in particular on the eigenvalue with the largest real part and the three eigenvector quantities associated with it, $|\psi^L|$, $|\psi^R|$, and $|\psi^L \psi^R|^{1/2}$. When $g = 0$ (no directional asymmetry), all three quantities are identically localized. However, when $g$ becomes nonzero, a hole opens up in the middle of the spectrum, converting the eigenvalues originally near the origin of the complex plane into a band gap with an expanding rim of weakly delocalized eigenstates. 
As $g$ increases (counterclockwise hopping bias), the eigenvalue with the largest real part initially stays the same, despite the changes in the middle of the spectrum, but the peaks of the localized right and left eigenvectors start to separate in opposite directions, and gradually widen as well. When $g$ reaches a high enough value such that the rim of the hole envelopes the principal eigenvalue, the separation between the left and right eigenvector peaks experiences a sudden jump, while the peak widths continue to gradually spread out. On the other hand, the profile of the product of the left and right eigenvectors (Fig.~\ref{fig:A2_g}b) does not change at all when $g$ initially increases from 0, provided that the location of the principal eigenvalue remains fixed in the complex plane. However, when the expanding rim reaches the principal eigenvalue, $|\psi^L \psi^R|^{1/2}$ suddenly becomes completely delocalized. 

The behavior of the three eigenvector quantities for the principal eigenvalue as a function of $g$, namely the gradual spreading of $|\psi^L|$ and $|\psi^R|$ as g increases, and the sudden complete delocalization of $|\psi^L \psi^R|^{1/2}$ when the eigenvalue is enveloped by the opening rim, is in fact experienced by all eigenvalues in the spectrum. A related phenomenon appears when IPR$_{R}$, IPR$_{L}$, and IPR$_{LR}$ are evaluated for all eigenvalues as a function of their distance from the expanding rim, at a fixed value of $g$. Fig.~\ref{fig:A2b_IPR_jump} plots IPR$_R$ and IPR$_{LR}$ as a function of the eigenvalue magnitude (distance away from the origin) for eigenvalues around the $45^o$ line in the complex plane, as indicated by the colored portions of the spectrum on the left. The data is extracted from 150 diagonalizations of the random sign hopping matrix $N=500$. Although these plots are rather noisy, IPR$_R$ decreases gradually as the eigenvalue gets closer to the rim, while IPR$_{LR}$ abruptly drops when the eigenvalue is right at the rim. 

Insight into these two behaviors follows from approximating the wavefunction magnitudes as wavepackets exponentially decreasing from their centers of localization~\cite{Hatano1997}. Then, using the similarity transformation in Ref.~\cite{Amir2016}, in a convenient continuum notation, we have,
\begin{eqnarray} \label{eq:waveg}
\left |\psi_n^R(x,g)\right| &\sim& e^{-\kappa_n | x-x_n| + gx} \\
\left|\psi_n^L(x,g)\right| &\sim& e^{-\kappa_n | x-x_n| - gx},
\end{eqnarray}
where $n$ labels the eigenfunction corresponding to the $n$-th eigenvalue, $x_n$ denotes the center of localization for $g=0$, and $\kappa_n$ is the Lyapunov exponent characterizing the exponential decay of the right and left eigenfunctions at $g=0$. An approximate inverse participation ratio (i.e. effective Lyaupnov exponent $\kappa^{eff}$) can then be calculated using Eq.~(\ref{eq:IPR}) and (\ref{eq:IPR_lr}),  
{\footnotesize{
\begin{eqnarray}
\text{IPR}_R(\lambda_n) &\approx& \frac{\int_{-\infty}^{+ \infty} dx |\psi_n^R(x,g)|^4}{\left(\int_{-\infty}^{+ \infty} dx |\psi_n^R(x,g)|^2 \right)^2}  \equiv \kappa_n^{eff, R}\\ 
\text{IPR}_{LR}(\lambda_n) &\approx&\frac{\int_{-\infty}^{+ \infty} dx |\psi_n^R(x,g)\psi_n^L(x,g)|^2}{\left(\int_{-\infty}^{+ \infty} dx |\psi_n^R(x,g)\psi_n^L(x,g)| \right)^2}  \equiv \kappa_n^{eff, LR},
\end{eqnarray}}} with the results
\begin{eqnarray} \label{eq:kappaeff}
\kappa^{eff, R}_n &\sim& \frac{\kappa_n^2 - g^2}{\kappa_n} \\ 
\kappa^{eff, LR}_n &\sim& \kappa_n. 
\end{eqnarray}
Thus, $\kappa^{eff, R}_n$ vanishes continuously as $g \rightarrow \kappa^-_n$, while $\kappa^{eff, LR}_n$ appears independent of $g$. Of course, we must remember that the approximate wavefunctions in Eq.~(\ref{eq:waveg}) become unormalizable when $g = \kappa_n$ and therefore $\kappa^{eff, LR}_n$ and IPR$_{LR}$ must cease to exist when $g = \kappa_n$. These rough arguments are consistent with the continuous evolution of IPR$_R$ as a function of g and the distance of the eigenvalue from the spectral rim (Fig.~\ref{fig:A2b_IPR_jump}a and ~\ref{fig:A2b_IPR_jump}a, respectively) and the sudden change in IPR$_{LR}$(g) and IPR$_{LR}(|\lambda|)$ (Fig.~\ref{fig:A2b_IPR_jump}b and ~\ref{fig:A2b_IPR_jump}b). Although the delocalization of the left and right eigenvectors is more gradual as $g \rightarrow \kappa_n^-$, they nevertheless mediate the sudden delocalization transition of the left-right inner product IPR$_{LR}$, when they leave the Hilbert space of localized states. 

\begin{figure}[htb]
\includegraphics[width=0.95\columnwidth]{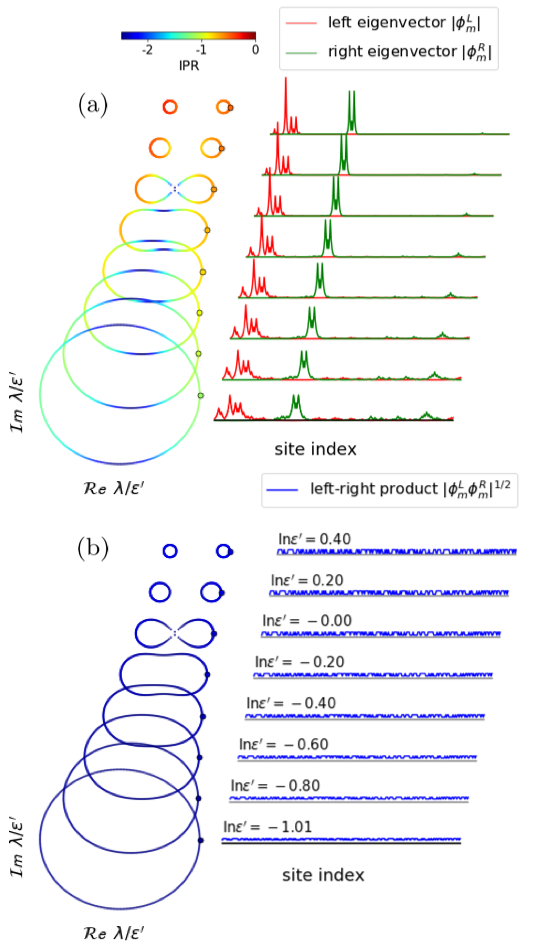}
\caption{\label{fig:A3_eps} \linespread{1}\selectfont{IPR$_R$ and $|\psi_R|$ and $|\psi_L|$ (top), and IPR$_{LR}$ and $|\psi_R \psi_L|^{1/2}$ (bottom) corresponding to the principal eigenvalue of one realization of the one-way hopping model and random sign connections (Eq.~(\ref{eq:Mprime})) with decreasing $\epsilon'$, the ratio between the disordered self-interaction strength and the strength of the one-way connection disorder. At large $\epsilon'$, the spatially separated left and right eigenvectors are fairly localized, spreading out more as $\epsilon'$ decreases. Meanwhile, the left-right inner product is delocalized at all values of $\epsilon'$ for the one-way model. }}
\end{figure}

\subsection{Consequences of directionality bias on dynamics}
In this final section, we comment briefly on the dynamics of a signal propagating in the one-dimensional one-way hopping model (large $g$ limit of Eq.~(\ref{eq:M})) with and without disordered self-interactions. 

\subsubsection{Signal current and eigenvalue velocity in the absence of diagonal disorder}

We first study the behavior of the mean and variance of a signal location in space as it propagates on a ring in the large-$g$ limit without onsite disorder (Eq.~(\ref{eq:Mprime}) with $\epsilon'  = 0$). We study a simple linear model like Eq.~(\ref{eq:M}), with $\bf M$ replaced by $\bf M'$ in Eq.~(\ref{eq:Mprime}). A more complete study would allow for multiple eigenvalues with positive real parts and include the effects of nonlinearities~\cite{Tanaka2018}. 

For the random sign hopping model studied in Sec.~\ref{sec:repulsion}, the sub-diagonal terms (clockwise connections) are to first order negligible in the large-$g$ limit. One can then ``gauge away'' the random signs on the counter-clockwise connections through a similarity transformation, and find the eigenvalues $\lambda_n$ and the strongly delocalized left and right eigenvectors $\psi_n^L$ and $\psi_n^R$ analytically (see also Sec.~\ref{sec:oriented} Eq.~(\ref{eq:wavenumber}))~\cite{Amir2016},
\begin{eqnarray}
\lambda_n(g) &=& e^{ik_n + g}\\
\psi_n^L & \sim & e^{-ik_nj}\\
\psi_n^R & \sim & e^{ik_nj}, \label{eq:psiR_stationary}
\end{eqnarray}
where
\begin{eqnarray}
k_n = \frac{2 \pi}{N}n, \quad n = 0, 1, \cdots, N-1. 
\end{eqnarray}

Let $\phi_0(x)$ denote a spatially localized signal at time $t=0$. Then, by expanding this initial state in a complex set of right eigenvectors, and then using the left eigenvectors to project out the expansion coefficients, we find the average position of the wave packet at time $t$,
{\footnotesize{
\begin{eqnarray}\label{eq:pos}
\langle x \rangle_t \sim \int_0^L dx~x~\sum_n \psi_n^R(x) e^{\lambda_n(g)t} \int_0^L dx'~\psi_n^L(x') \phi_0(x'). 
\end{eqnarray}}}Since the integral over $x'$ does not depend on $x$, we denote 
\begin{eqnarray}
f_n \equiv \int_0^L dx'~\psi_n^L(x') \phi_0(x')
\end{eqnarray}
in all subsequent equations. Normalizing Eq.~(\ref{eq:pos}) then gives the following,
\begin{eqnarray}
\langle x \rangle_t &=& \frac{\int_0^L dx~x~\sum_{n=0}^{N-1}~ \psi_n^R(x) e^{\lambda_n(g)t} f_n}{\int_0^L dx~\sum_{n=0}^{N-1}~ \psi_n^R(x) e^{\lambda_n(g)t} f_n} .
\end{eqnarray}
Upon integrating by parts, we find
\begin{eqnarray} \label{eq:xt}
\langle x \rangle_t - \langle x \rangle_0 = - \lambda_0(g) t = -e^g t = -t \frac{d\lambda_{0}}{dg},
\end{eqnarray}
where $\lambda_0$ denotes the ``ground state'' eigenvalue corresponding to the lowest wave number $k_0$, which for this problem is the eigenvalue with the largest real part. The minus sign is present in Eq.~(\ref{eq:xt}) because the hopping is biased in the counterclockwise direction.

The time evolution of the second moment associated with this initial condition is found from
\begin{eqnarray}
\langle x^2 \rangle_t &=& \frac{\int_0^L dx~x^2~\sum_{n=0}^{N-1}~ \psi_n^R(x) e^{\lambda_n(g)t} f_n}{\int_0^L dx~\sum_{n=0}^{N-1}~ \psi_n^R(x) e^{\lambda_n(g)t} f_n} ,
\end{eqnarray}
which leads to
\begin{eqnarray}
\langle x^2 \rangle_t - \langle x^2 \rangle_0 = t \lambda_0(g) \left(1 + \lambda_0(g) t \right).
\end{eqnarray}
After incorporating Eq.~(\ref{eq:xt}), we find that the variance describing the spreading of this wave packet grows linearly in time
\begin{eqnarray}
\left( \langle x^2 \rangle_t - \langle x \rangle_t^2 \right)  - \left(\langle x^2\rangle_0 - \langle x \rangle_0^2 \right) = t \frac{d\lambda_0}{dg}. 
\end{eqnarray}
One can also obtain the same behavior by directly calculating the signal at time $t$, $\phi_t(x)$, starting with a Gaussian initial condition $\phi_0(x) \sim \exp(-x^2/2a)$ at $t=0$ . 

To summarize, a signal propagating on the ring with one-way random sign hopping and no onsite disorder travels with a constant speed $e^g$, and has a standard deviation that increases as $\sqrt{t}e^{g/2}$. In the long time limit, the signal stops spreading when it covers the entire ring and converges to a flat stationary  state, given by $\psi^R_0$ in Eq.~(\ref{eq:psiR_stationary}). 

\subsubsection{Localized response mediated by spatially separated left and right eigenvectors with diagonal disorder}

In the previous section, we saw that regardless of the initial condition (excitation signal), the one-way hopping model without onsite disorder allows a signal to propagate and spread out on a ring of connections as a function of time, eventually converging to a stationary delocalized state. However, this behavior is dominated by delocalized eigenvectors. When self-interaction disorder is incorporated, the dynamics is quite different, because a large portion of the spectrum exhibits localized eigenvectors in the presence of onsite disorder, even for models with one-way connections.

In addition to localization effects, the response of hopping models with directional bias has an additional interesting property that is not present in non-biased hopping models. Because of the separation of the left and right eigenvectors when there is nonzero directional bias, a localized response (at the peak of the right eigenvector) can be triggered at a considerable distance away from the location of the excitation signal (at the peak of the left eigenvector) via the propagator of the dynamical models associated with matrices $\bf{M'}$ studied in Sec.~\ref{sec:oriented},
\begin{eqnarray}
G(x,x',t) = \sum_{n = 0}^{N-1} \psi_n^{R}(x) \psi_n^{L}(x') e^{\lambda_{n} t}.
\end{eqnarray}

Fig.~\ref{fig:A3_eps}a shows the left and right eigenvectors corresponding to the principal eigenvalue, for one realization of the random sign one-way hopping model (Eq.~(\ref{eq:Mprime})). Here, the tuning parameter is $\epsilon'$, the ratio between the disordered self-interaction strength and the strength of the one-way connection disorder. Even at large $\epsilon'$, although the onsite disorder essentially pins down the signal such that it does not travel or spread, the system can nevertheless sense the excitation signal and respond at distances on the order of the system size. Interestingly, as shown in Fig.~\ref{fig:A3_eps}b, the product of the left and right eigenvector $|\psi^R \psi^L|^{1/2}$ is completely delocalized for all eigenvalues in the spectra at all values of $\epsilon'$. 

\bibliography{sparse_RMT}

\end{document}